\documentclass[preprint,showpacs,preprintnumbers,amsmath,amssymb]{revtex4}

\usepackage{graphicx}
\usepackage{dcolumn}
\usepackage{bm}
\usepackage{appendix}
\newcommand {\apgt} {\ {\raise-.5ex\hbox{$\buildrel>\over\sim$}}\ }
\newcommand {\aplt} {\ {\raise-.5ex\hbox{$\buildrel<\over\sim$}}\ }

\begin{document}

\title{Scaling laws for Shor's algorithm with a banded quantum Fourier transform} 

\author{Y. S. Nam and R. Bl\"umel}  
\affiliation{Department of Physics, Wesleyan University, 
Middletown, Connecticut 06459-0155}
  
\date{\today}

\begin{abstract} 
We investigate the performance of a streamlined 
version of Shor's algorithm in which the 
quantum Fourier transform is 
replaced by a banded version that for each 
qubit retains only coupling to its $b$ 
nearest neighbors. Defining the 
performance $P(n,b)$ of the 
$n$-qubit 
algorithm for bandwidth $b$
as the ratio of the success rates of 
Shor's algorithm equipped with the 
banded and the full bandwidth ($b=n-1$) 
versions of the quantum Fourier transform, 
our numerical simulations show that
$P(n,b) \approx \exp[-\varphi_{max}^2 (n,b)/100]$ for
$n < n_t(b)$ (non-exponential regime) and
$P(n,b) \approx 2^{-\xi_b (n-8)}$ 
for $n>n_t(b)$ (exponential regime), where
$n_{t}(b)$, the location of the transition, is
approximately given by $n_{t}(b)\approx b+5.9 + \sqrt{7.7(b+2)-47}$
for $b\gtrsim 8$,
$\varphi_{max} (n,b) = 2\pi[2^{-b-1} (n-b-2) + 2^{-n}]$,
and
$\xi_b\approx 1.1 \times 2^{-2b}$.
Analytically we obtain
$P(n,b) \approx \exp[-\varphi_{max}^2 (n,b)/64]$ for $n<n_t(b)$
and $P(n,b) \approx 2^{-\xi_b^{(a)} n}$ for $n>n_t(b)$, where
$\xi_{b}^{(a)} \approx \frac{\pi^2}{12 \ln(2)} \times 2^{-2b} \approx 1.19 \times 2^{-2b}$.
Thus, our analytical results predict the $\varphi_{max}^2$ scaling
($n<n_t$) and the $2^{-2b}$ scaling ($n>n_t$) of the 
data perfectly. In addition, in the large-$n$ regime, the
prefactor in $\xi_b^{(a)}$ is close to the results of
our numerical simulations and, in the low-$n$ regime, the
numerical scaling factor in our analytical result is within
a factor $2$ of its numerical value.
As an example we show that $b=8$ is sufficient
for factoring RSA-2048 with a 95\% success rate.
  
\end{abstract}

                       


\maketitle
%
%
\section{Introduction}
While the art of integer factoring lay dormant, literally for millennia, 
and not much progress beyond the crudest methods, such as 
trial division and looking for differences of squares, had been made 
\cite{Pomerance}, 
the advent of 
the widely used RSA cryptosystem \cite{RSA1} 
has recently propelled the factoring of large integers from the 
arcane recesses of an ancient mathematical discipline into the lime light 
of contemporary physics and mathematics. 
The reason is that a powerful factoring algorithm may be used in 
a frontal attack on the RSA cryptosystem, and, if successful, immediately 
reveals untold scores of government, military, and 
financial secrets \cite{RSA2,RSA3}. No wonder then, that the first substantial 
breakthrough in factoring in centuries, the quadratic number sieve 
\cite{QuadSieve,Pomerance}, 
occurred shortly after the 
initial publication of the RSA method \cite{RSA1}. 
Using the quadratic number sieve, RSA keys with up to 
100 decimal digits can now routinely be cracked \cite{QScrack} and 
are not safe any more. In 1993, the 
general number field sieve \cite{GNFS} 
added even more power to factoring 
attacks on RSA and was used successfully to factor the 
RSA challenge number RSA-768 (232 decimal digits) \cite{Kleinjung}, which 
prompted the US National Institute of Standards and Technology 
(NIST) to recommend retirement of all RSA keys with 1024 binary 
digits or less \cite{NIST}. However, no matter how powerful these modern 
factoring algorithms are, they are based on classical 
computing algorithms, executed on classical computers and 
without further improvements 
will never be able to crack an RSA key consisting of 
5000 decimal digits or more (see Sec.~\ref{SectionDC}). 
But not only classical computing profited from the advent of 
the RSA crypto-system, so did quantum computing \cite{NC}. 
In 1994, Shor demonstrated that a certain quantum algorithm
executed on a quantum computer is 
exponentially more powerful than any currently known classical 
factoring scheme and poses a 
real threat to RSA-encrypted data \cite{Shor1}. 
Since its inception in 1994, Shor's algorithm has maintained 
its status as the gold standard in quantum computing, and 
progress in quantum computer implementation is frequently 
measured in terms of the size of semiprimes that
a given quantum computer can factor \cite{ERShor1,ERShor5}. 
While, compared with classical factoring algorithms, 
Shor's algorithm is tremendously more powerful, it should not 
come as a surprise that in order to break 
currently employed RSA keys, 
an enormous number of quantum operations still need to be performed. 
Therefore, any advance in streamlining practical implementations 
of Shor's algorithm are welcome that result in reducing the 
number of required quantum operations. 
A central component of Shor's algorithm is a 
quantum Fourier Transform \cite{NC} and 
our paper focuses on how to perform this part of 
Shor's algorithm with the least number of quantum gates and 
gate operations that still guarantee acceptable performance 
of the algorithm.

Our paper is organized in the following way. In Sec.~\ref{SectionShor} we present Shor's algorithm. This section also serves to 
introduce the basic notation and explains the central position of the quantum Fourier transform
in Shor's algorithm. While the original version of Shor's algorithm \cite{Shor1} is formulated 
with the help of a full implementation of the quantum Fourier transform, it turns out that 
a reduced, approximate version of the quantum Fourier transform, the banded quantum 
Fourier transform \cite{Copper,FH,Nam}, yields surprisingly good results when used in conjunction with Shor's algorithm. The banded quantum Fourier transform is introduced and discussed 
in Sec.~\ref{SectionBQFT}. In order to assess the influence of the banded quantum Fourier 
transform on the performance of Shor's algorithm, we need an objective performance measure. 
Our performance measure is defined in Sec.~\ref{SectionPM}. In Sec.~\ref{SectionNR}, based on 
the performance measure defined in Sec.~\ref{SectionPM}, we investigate numerically the 
performance of a quantum computer for various bandwidths $b$ as a function of the number of
qubits $n$. We find that for fixed $b$ the quantum computer exhibits two qualitatively different
regimes, exponential for large $n$ and non-exponential for small $n$.
We also find that relatively small $b \lesssim 10$ 
are already sufficient for excellent quantum
computer performance, even for $n$ so large as to be interesting for the factoring of 
semiprimes $N$ of practical interest. These numerical findings are then investigated analytically 
in Sec.~\ref{SectionAR}. In Sec.~\ref{SectionARA}, we show an important property of the
performance measure, i.e. approximate separability,
which allows us to analyze analytically the large-$n$ 
behavior (Sec.~\ref{SectionARB}) and the small-$n$ behavior (Sec.~\ref{SectionARC}) of the 
numerical data presented in Sec.~\ref{SectionNR}. In particular, we are able to predict analytically
the scaling functions of the data in the large-$n$ and small-$n$ regimes.
In Sec.~\ref{SectionFH} we compare our work with the related pioneering work of Fowler
and Hollenberg \cite{FH}. While the final results are similar, our approach differs
substantially from the approach in \cite{FH}. Factoring actual semiprimes, our approach
is more realistic than the approach taken in \cite{FH} and may serve to check the results
reported in \cite{FH}. In addition, we report a host of new results.
In Sec.~\ref{SectionDC} we discuss our results and conclude the paper 
in Sec.~\ref{SectionSC}. In order not to break the 
flow of exposition in the main text of our paper, 
some technical material is relegated to three 
appendices. In Appendix~\ref{AppendixA} we prove existence and uniqueness of 
an order-2 element for any semiprime $N$. In Appendix~\ref{AppendixB} we compute
an analytical bound for the maximal possible order $\omega$ of a given semiprime $N$.
In Appendix~\ref{AppendixC}, we provide an auxiliary result on the distribution of 
an inverse factor of $\omega$, needed for one of our analytical results reported in Sec.~\ref{SectionAR}.


\section{Shor's Algorithm} 
\label{SectionShor}
Progress in quantum computing happens in fits and starts. 
Periods of stagnation and pessimism are followed by 
unexpected breakthroughs and optimism. Shor's algorithm 
is a case in point. Following a lull in quantum 
computing during which the only known quantum algorithms 
were of an ``academic'' nature, Shor's 
algorithm, the first ``useful'' quantum algorithm, 
instantly revived the field when it burst on the 
scene, quite unexpectedly, in 1994 \cite{Shor1}. 
Shor's algorithm is 
quantum mechanics' answer to a task that is 
hard or impossible to perform on 
any classical computer: factoring large 
semiprimes $N$. To accomplish this task, 
Shor's algorithm makes use of the entire 
palette of quantum effects 
that result in an exponential 
speed-up of the quantum algorithm with respect 
to any currently known classical factoring algorithm: 
superposition, interference, and entanglement. 
Shor's algorithm is based on Miller's algorithm \cite{Blu}, 
a classical factoring algorithm. 
Miller's algorithm determines the factors of a 
semiprime $N=pq$, where $p\neq q$ are prime, 
according to the following procedure. 
First, we choose a positive integer 
$1 < x < N$, called the seed, relatively prime 
to $N$, i.e. ${\rm gcd}(x,N)=1$, where 
${\rm gcd}$ denotes the greatest common divisor. 
Then, we determine the smallest positive integer 
$\omega$, called the order of $x$, such that 
\begin{equation}
x^{\omega} \mod N = 1.
\label{Shor1}
\end{equation}
For Miller's algorithm to work, we require 
(i) that $\omega$ is even and (ii) that 
$(x^{\omega/2}+1)\mod N\neq 0$. Both conditions 
need to be fulfilled. If even one is not fulfilled, 
we need to choose 
another $x$ and try again. There is a high probability that 
this will succeed after only a few trials 
\cite{NC,FH,Mermin}. 
Having found 
a seed $x$ satisfying both conditions, we 
write (\ref{Shor1}) in the form 
\begin{equation}
[(x^{\omega/2} - 1) (x^{\omega/2} + 1) ] \mod N = 0, 
\label{Shor3}
\end{equation}
which implies that $N$ divides the product on the left-hand side of 
(\ref{Shor3}). This might be accomplished if $N$ divides 
$x^{\omega/2}-1$, which implies $x^{\omega/2}\mod N=1$. 
This, however, is impossible, because $\omega/2<\omega$ and 
$\omega$,  according to (\ref{Shor1}), 
is the smallest such exponent. Another hypothetical possibility is that 
$N$ divides the second factor in (\ref{Shor3}). 
This, however, is excluded according to condition (ii). 
The only remaining possibility is that $p$ divides one of the 
factors in (\ref{Shor3}) and $q$ divides the other. Appropriately 
naming the factors of $N$, we have 
\begin{equation}
p = {\rm gcd}(x^{\omega/2} - 1,N),\ \ \ 
q = {\rm gcd}(x^{\omega/2} + 1,N), 
\label{Shor4}
\end{equation}
and the factoring problem is solved. So, if Miller's classical algorithm does the job, 
why do we need Shor's quantum algorithm? The answer is that 
finding the order $\omega$ on a classical computer is an algorithmically hard problem that,
for a generic seed $x$, is impossible to perform on a classical computer within a reasonable 
execution time for semiprimes $N$ with more than 5000 digits (see Sec.~\ref{SectionDC}). 
This is where Shor's algorithm comes in. Using a quantum Fourier transform to 
find the order $\omega$, Shor's algorithm makes order-finding tractable on 
a quantum computer. This is how it works. 
 
First, we define the function 
\begin{equation}
f(k) = x^k \mod N, 
\label{Shor5}
\end{equation}
where $k$ is an integer with $k\geq 0$. Since 
$f(k+\omega)=f(k)$, the function $f$ turns order finding 
into period finding. Since periods may be found 
by a Fourier transform, the central idea of Shor's algorithm 
is to use a quantum Fourier transform to determine $\omega$. 
To implement this idea \cite{Shor1,NC,Mermin,Blu}, 
we work with a quantum computer consisting of two 
quantum registers, register $I$ and register $II$. We assume 
that both registers consist of $n$ qubits. 
In order to reliably determine $\omega$ for given $N$,
care must be taken to choose $n$ at least twice 
as large as the number of binary digits of $N$ \cite{NC,Mermin}.
We strictly observe this requirement in Sec.~\ref{SectionNR} [see (\ref{NUMRES1})],
where we present our numerical work.
We start by 
initializing both registers to 0 such that the initial 
state of the quantum computer is 
\begin{equation}
|\psi\rangle = |0,\ldots,0\rangle_{I} \,  |0,\ldots,0\rangle_{II}. 
\label{Shor6}
\end{equation}
Next, we initialize register $I$ with a superposition of 
all integers from 0 to $2^n-1$ by applying a single-qubit 
Hadamard transform \cite{NC} to each 
of the $n$ qubits of register $I$, resulting in the state 
\begin{equation}
 |\psi\rangle = \frac{1}{\sqrt{2^n}}\, 
 \sum_{k=0}^{2^n-1}\, |k\rangle_{I} \, |0,\ldots,0\rangle_{II}, 
 \label{Shor7}
 \end{equation}
where we introduced an intuitive equivalence, whereby 
an integer $k\geq 0$ is mapped onto the 
$n$ qubits of a register according to the binary digits of $k$. 
Now, we make use of the function $f$ defined in (\ref{Shor5}) 
to fill register $II$ with the 
$f$-images of register $I$. This results in the computer state 
\begin{equation}
 |\psi\rangle = \frac{1}{\sqrt{2^n}}\, 
 \sum_{k=0}^{2^n-1}\, |k\rangle_{I} \, | f(k) \rangle_{II}. 
 \label{Shor8}
 \end{equation}
This step entangles registers $I$ and $II$.
The function $f$ induces equivalence classes 
\begin{equation}
[s_0] = \{ s_0+k\omega,\ \ \ 0\leq k\leq K(s_0)-1 \} 
\label{Shor9}
\end{equation}
on $\{0,\ldots,2^n-1\}$ with representatives 
$0\leq s_0\leq \omega - 1$, where $K(s_0)$ is the smallest 
integer with 
$s_0+K(s_0)\omega\geq 2^n$. In other words, 
$K(s_0)$ is the number of elements in the 
equivalence class $[s_0]$. Since the range of $s$ values is $2^n$
and the spacing is $\omega$, we obtain, approximately,
\begin{equation}
\label{K}
K(s_0) \approx \frac{2^n}{\omega}.
\end{equation}
Because of the periodicity of 
$f$, each member of $[s_0]$ is mapped onto $f(s_0)$. 
Therefore, if a measurement of register $II$ collapses this 
register into the state $|f(s_0)\rangle_{II}$, 
the quantum computer is in the state 
\begin{equation}
|\psi_{i}\rangle = \frac{1}{\sqrt{K(s_0)}}\, 
\sum_{k=0}^{K(s_0)-1} |s_0+k\omega\rangle_I \, |f(s_0)\rangle_{II}. 
\label{Shor10}
\end{equation}
We may now apply a quantum Fourier transform 
\begin{equation}
\hat U^{(QFT)} = \frac{1}{\sqrt{2^n}}\, \sum_{k,l=0}^{2^n-1}\, 
|l\rangle \, \exp(2\pi i l k / 2^n) \, \langle k| 
\label{Shor11}
\end{equation}
to register $I$ of $|\psi_i \rangle$ to obtain 
\begin{equation}
|\psi_{f}\rangle = \frac{1}{\sqrt{K(s_0)2^n}}\, \sum_{k=0}^{K(s_0)-1} 
\sum_{l=0}^{2^n-1} \, \exp[2\pi i l (s_0+k\omega)/2^n] \, 
| l \rangle_I \, |f(s_0)\rangle_{II}. 
\label{Shor12}
\end{equation}
A measurement of register $I$ then collapses $|\psi_{f}\rangle$ 
into $| l\rangle$ with probability 
\begin{align}
\tilde P(n,l,\omega) &= \frac{1}{2^n K} \, \left| \sum_{k=0}^{K-1} 
\exp(2\pi i l k \omega / 2^n) \right| ^2 
\nonumber \\ 
&= \frac{\sin^2(K\pi \omega l/2^n)} {2^n K 
\sin^2(\pi\omega l/2^n)}, 
\label{Shor13}
\end{align}
where here and in the following we suppressed the argument $s_{0}$ of $K$.
Apparently, $\tilde P(n,l,\omega)$ is sharply peaked at $l$ values for which 
$\omega l / 2^n$ is close to an integer. As a consequence, these $l$ values 
will appear as a result of measurement with a high probability. 
Subsequent 
analysis of the measured peak location on 
a classical computer then reveals the factors of $N$ with high probability 
\cite{NC}. This step is called classical 
post processing \cite{NC,Mermin}. 
Equation (\ref{Shor13}) is the starting point of our 
analysis of the performance of Shor's algorithm with a banded 
quantum Fourier transform in Sec.~\ref{SectionPM}. 
  
Several experimental demonstrations of Shor's algorithm 
have been published \cite{ERShor1,ERShor2,ERShor3,ERShor4,ERShor5}. 
Since it is exceedingly difficult to experimentally control more than 
a handful of qubits, the numbers 
$N$ factored in these experiments are very small, currently not exceeding 
$N=21$ \cite{ERShor5}. 
Therefore, reaching higher $N$ is facilitated by reducing 
the requirements to run 
Shor's algorithm on a quantum computer. One such optimization is 
the use of an approximate, banded quantum Fourier transform 
\cite{Copper} instead of the the full quantum 
Fourier transform (\ref{Shor11}). Further optimization is possible 
by using a banded version of the semi-classical quantum Fourier transform 
\cite{GN} defined in the following section. 
              
 
\section{Banded Quantum Fourier Transform}
\label{SectionBQFT}
\begin{figure}
\centering
\includegraphics[scale=.4,angle=-90]{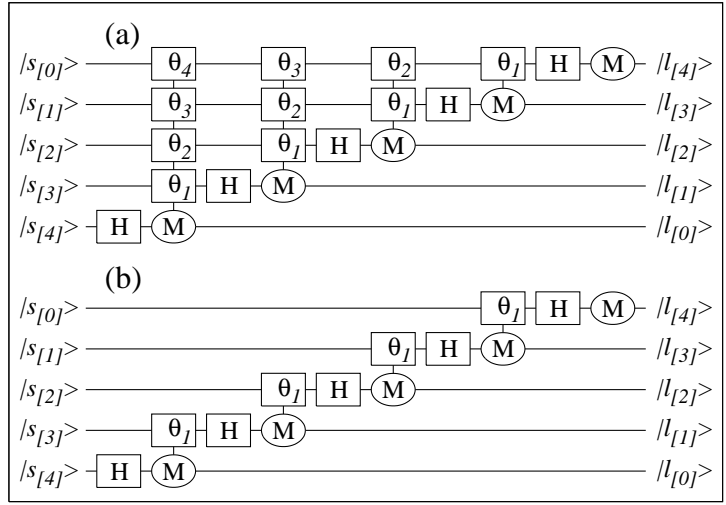}
\caption{\label{fig1} Logic circuit of a five-qubit implementation of 
the single-qubit realization 
of the quantum Fourier transform \cite{GN}. 
(a) Full implementation (bandwidth $b$ 
= 4); (b) truncated 
implementation (bandwidth $b$ 
= 1). H, $\theta$, and M denote Hadamard, 
single-qubit conditional rotation, and measurement gates, 
respectively.        
}
\end{figure}

A direct circuit implementation of the Fourier transform defined in (\ref{Shor11})
requires 
$n(n+1)/2$ two-qubit quantum gates \cite{NC}. 
In \cite{GN}, it was shown 
that, when followed by measurements, as required by Shor's algorithm, 
an equivalent quantum circuit, consisting exclusively of single-qubit gates, is 
exactly equivalent to the two-qubit realization of the quantum Fourier 
transform. Figure~\ref{fig1} (a) 
illustrates this single-qubit realization of the quantum 
Fourier transform for the special case of five qubits
(we classify the conditional rotation gates $\theta$ in Fig.~\ref{fig1} as
single-qubit gates since they are controlled by classical input and act
coherently only on a single qubit).
This circuit still 
requires $\sim n^2$ gate operations, but since they are performed by single-qubit 
gates, experimental implementation of this single-qubit circuit is considerably 
simpler. In contrast to the full two-qubit implementation of the 
quantum Fourier transform, where the measurements may 
occur simultaneously at the end of the quantum computation, 
the measurements in the single-qubit version of the quantum 
Fourier transform [denoted by the M gates in 
Fig.~\ref{fig1}~(a)] occur sequentially and 
their (classical) measurement results are used to 
control the phase rotation gates $\theta$. 
As first pointed out by 
Coppersmith \cite{Copper}, even this quantum circuit may 
still be optimized by working with an approximate, banded 
quantum Fourier transform as illustrated in 
Fig.~\ref{fig1}~(b). 
 
The banded quantum Fourier transform $\hat{U}_{b}^{(QFT)}$ [see Fig.~\ref{fig1}~(b)] 
is obtained from the full implementation of the single-qubit 
quantum Fourier transform [see Fig.~\ref{fig1}~(a)] 
by retaining only the coupling to $b$ nearest neighbors of a given 
qubit. As illustrated in Fig.~\ref{fig1}~(b)
for the case $b=1$, this results 
in a banded structure of the corresponding quantum circuit 
\cite{Nam}. 
The name is also justified on theoretical grounds since the 
unitary matrix representing the circuit shown in 
Fig.~\ref{fig1}~(b) has a banded structure 
\cite{PVTF}. 
The banded 
quantum Fourier transform of bandwidth $b$ is the basis of 
our work presented in the following sections.  
%
 
\section{Performance Measure}
\label{SectionPM} 

The key idea of Shor's algorithm is to use superposition 
and entanglement to steer the quantum probability
into qubits that correspond to numbers encoded in binary
form, which will then, as a result of classical post-processing, 
reveal the factors of $N$. Our first task, therefore, is to locate
the useful peaks after the quantum Fourier transform is performed.
In order to define
our performance measure, we are interested in how sharp these
peaks are in $l$. For this purpose, we notice that $\tilde{P}(n,l,\omega)$
[see (\ref{Shor13})] (up to a factor) is of the form
\begin{equation}
f(z) = \frac{\sin^{2}(Kz)}{\sin^{2}(z)},
\end{equation}
where $K$ is a large integer, $z$ is a real number, and $f(z)$ is sharply
peaked at integer multiples of $\pi$. Since 
the shape of $f(z)$ is the same for $z$ in the vicinity of each peak,
it suffices to investigate the peak at $z=0$ to determine 
the width of all the other peaks of $f(z)$. We define the half width $\Delta z$ of
$f(z)$ by requiring
\begin{equation}
\label{halfwidth}
f(\Delta z) = \frac{1}{2}.
\end{equation}
Inspired by a second-order Taylor-series expansion of (\ref{halfwidth}), 
we obtain the heuristic formula
\begin{equation}
\Delta z \approx \frac{1.39}{K},
\end{equation}
which, for $K > 10$, satisfies (\ref{halfwidth}) to better than $10^{-3}$.
Applied to $\tilde{P}(n,l,\omega)$ in (\ref{Shor13}), we have 
\begin{equation}
z = \frac{\pi \omega l}{2^{n}},
\end{equation}
and, therefore,
\begin{equation}
\Delta z = \frac{\pi \omega}{2^{n}} \Delta l \approx \frac{1.39}{K},
\end{equation}
from which we obtain
\begin{equation}
\label{peakwidth}
\Delta l \approx \bigg( \frac{2^{n}}{\omega K} \bigg) \bigg( \frac{1.39}{\pi} \bigg) \approx 0.44,
\end{equation}
where we used (\ref{K}).
This result shows that the full width at half maximum of the $l$-peaks is 
only about one state and that this width is ``universal'' in the sense that 
it is independent of $K$, $\omega$, and $n$.

Since a peak in $\tilde{P}(n,l,\omega)$ 
occurs whenever $\omega l / 2^{n}$ is close to an integer, we define the
$l$-integer closest to peak number $j$ according to:
\begin{equation}
\label{peakloc}
l_{j} = \bigg( \frac{2^{n}}{\omega}\bigg) j + \beta_{j}, \ \ \ j = 0,1, \dots , \omega-1,
\end{equation}
where $\beta_{j}$, a rational number, ranges between $-1/2$ and $1/2$. 
Since the peaks in $\tilde{P}(n,l,\omega)$ are universal in the above sense
and contain basically only a single state, namely $l_{j}$ defined in (\ref{peakloc}), 
we use
\begin{equation}
\tilde{P}(n,l_{j},\omega) \equiv \tilde{P}_{j}(n,\omega)
\end{equation}
as the basis for our performance measure.

\begin{figure}
\centering
\includegraphics[scale=1]{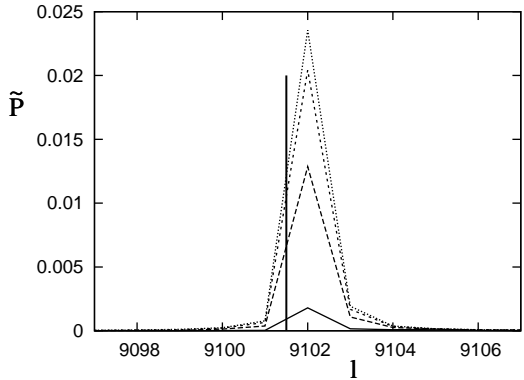}
\caption{\label{widthfigure} 
Shape of a Fourier peak in $l$ as a function of $b$ for
the semiprime $N = 247$ and order $\omega = 36$.
Shown are the peaks for different bandwidths $b=1$ (solid),
$b=2$ (long-dashed), $b=3$ (short-dashed), and $b=10$ (dotted).
The vertical solid line is located at $l=9101.5$ .
  }
\end{figure}

Although the width of the peaks of $\tilde{P}(n,l,\omega)$ is narrow, 
according to (\ref{peakwidth}) of the order of a single state, and 
although $|l_{j}\rangle$ carries most of the probability in peak number 
$j$ of $\tilde{P}(n,l,\omega)$ (approximately $77\%$ on average), there 
are nevertheless several states $|l\rangle$ inside of peak number $j$ 
that occur with a small but still appreciable probability in a measurement 
of $|\psi_{f}\rangle$ in (\ref{Shor12}). These states are also useful 
for factoring during classical post-processing (see Sec.~\ref{SectionShor}
and \cite{NC,Mermin}), 
and the question arises if these states should be included in the performance 
measure. Indeed, instead of determining the performance of Shor's algorithm 
on the basis of the single state $|l_{j}\rangle$, Fowler and Hollenberg \cite{FH}, e.g., 
base their performance measure on the two closest states to the peaks in 
$\tilde{P}(n,l,\omega)$. We found that including more states in the performance measure 
is not necessary, since the width of the Fourier peaks in $l$ is 
independent of the bandwidth $b$. At first glance this is 
surprising, since intuitively, we would think that the 
quality of the quantum Fourier transform should deteriorate 
with decreasing bandwidth $b$, possibly accompanied 
by a broadening of the Fourier peaks in $l$. That this 
is not so, and that the widths of the Fourier peaks are indeed 
independent of $b$, is demonstrated in Fig.~\ref{widthfigure} for 
the case $N=247$ for $b=1,2,3,10$. 
Independent of $b$, the vertical line in the 
figure cuts each Fourier peak at approximately its 
midpoint, thus demonstrating that the widths of 
the Fourier peaks in $l$ are indeed independent of $b$. 
Thus, upon a change in $b$, all $l$ states under a 
Fourier peak respond in unison to the change in $b$. 
Therefore, a single $l$ state, such as $l_j$, is an 
excellent representative of all the $l$ states in 
its immediate vicinity. 

Defining $\tilde{P}_j(n,b,\omega) = \tilde{P}(n,l_j,b,\omega)$ as the probability of obtaining $|l_{j}\rangle$ 
in a measurement of $|\psi_{f}\rangle$ if instead of the full quantum Fourier transform
(\ref{Shor11}) the banded quantum Fourier transform (see Sec.~\ref{SectionBQFT}) 
is used and taking into account that the widths of the peaks in $\tilde{P}_j(n,b,\omega)$ 
do not change as $b$ is varied,
we use the ratio of the total probability of collapse 
into one of the states $|l_{j}\rangle$, given the bandwidth $b$, to that of the full bandwidth 
$b = n-1$, to capture the overall probability of obtaining the useful $|l\rangle$ states in
the vicinity of $|l_j\rangle$. Thus, the normalized ratio is of the form
\begin{equation}
\label{measure}
P(n,b,\omega) = \tilde{P}(n,b,\omega)/\tilde{P}(n,b=n-1,\omega),
\end{equation}
where
\begin{equation}
\label{rawmeasure}
\tilde{P}(n,b,\omega) = \sum_{j=0}^{\omega-1} \tilde{P}_{j}(n,b,\omega)
\end{equation}
and $\tilde{P}(n,b=n-1,\omega)$ is the probability of collapsing into any
one of the set of useful states $|l_j \rangle$ as a result of measuring $|\psi_f \rangle$, 
where $|\psi_{f}\rangle$ is generated 
from $|\psi_{i}\rangle$ by application of the full quantum Fourier transform 
$\hat{U}^{(QFT)}$ defined in (\ref{Shor11}). We use $P(n,b,\omega)$,
defined in (\ref{measure}), as our performance measure throughout this paper.

Next, we derive an analytical expression for $\tilde{P}_j(n,b,\omega)$, valid 
for any bandwidth $0 \leq b \leq n-1$, that can be used in our performance measure 
(\ref{measure}). In order to find $\tilde{P}_{j}(n,b,\omega)$ we need to descend to the 
qubit-by-qubit level, since the bandwidth $b$ in $\hat{U}_{b}^{(QFT)}$ refers to 
inter-qubit spacing on the qubit level in the circuit diagram of $\hat{U}_{b}^{(QFT)}$ 
[see Fig. \ref{fig1}~(b)]. We start with a representation of the quantum Fourier transform 
in bit-notation
\begin{align}
\hat{U}^{(QFT)}|s\rangle &= \frac{1}{\sqrt{2^n}} \sum_{l=0}^{2^n-1} 
e^{\frac{2\pi isl}{2^{n}}} |l\rangle \cr 
&= \frac{1}{\sqrt{2^{n}}} \prod_{m=0}^{n-1} \sum_{l_{[n-m-1]}=0}^{1} 
e^{2\pi i(.s_{[m]}s_{[m-1]}\dots s_{[0]})l_{[n-m-1]}}|l_{[n-m-1]}\rangle,
\end{align}
where $s_{[\nu]}(l_{[\nu]})$ indicates the $\nu$th binary digit of $s$ 
($\nu$th binary digit of $l$) and 
\begin{equation}
\big(.s_{[m]}s_{[m-1]}\dots s_{[0]}\big) = \sum_{\nu=0}^{m} s_{[\nu]} 2^{-(m-\nu+1)}.
\end{equation}
For bandwidth $b$, $\hat{U}_{b}^{(QFT)}|s\rangle$ then becomes 
\begin{equation}
\hat{U}_{b}^{(QFT)}|s\rangle = \frac{1}{\sqrt{2^{n}}} \prod_{m=0}^{n-1} 
\sum_{l_{[n-m-1]}=0}^{1} e^{2\pi i[(.s_{[m]}s_{[m-1]}\dots s_{[0]}) - (.00\dots0s_{[m-b-1]}\dots s_{[0]})
]l_{[n-m-1]}}|l_{[n-m-1]}\rangle.
\end{equation}
We may also write
\begin{equation}
\label{UQFT_b}
\hat{U}_{b}^{(QFT)}|s\rangle = \sum_{l=0}^{2^n-1} B(s,l)|l\rangle,
\end{equation}
where
\begin{equation}
B(s,l) = \frac{1}{\sqrt{2^n}} \exp{\Bigg\{ 2\pi i \sum_{m=0}^{n-1} \big[
\Lambda_{m,0}(s) - \Lambda_{m,b+1}(s) \big] l_{[n-m-1]}\Bigg\}},
\end{equation}
and
\begin{equation}
\Lambda_{m,\lambda}(s) = \big(.00\dots0s_{[m-\lambda]}s_{[m-\lambda-1]}\dots s_{[0]}\big),
\end{equation}
i.e. $\lambda$ zeros are following the binary point. Defining
\begin{equation}
S_{\lambda}(s,l) = \sum_{m=0}^{n-1} \Lambda_{m,\lambda}(s) l_{[n-m-1]} ,
\end{equation}
we may express $B(s,l)$ in the form
\begin{equation}
\label{B(s,l)}
B(s,l) = \frac{1}{2^{n/2}} \exp{\Big\{ 2\pi i[S_{0}(s,l)-S_{b+1}(s,l)] \Big\}}.
\end{equation}
Sorting indices, $S_{\lambda}(s,l)$ may be written in the form
\begin{equation}
\label{S(s,l)}
S_{\lambda}(s,l) = \frac{1}{2} \sum_{m=\lambda}^{n-1} \sum_{\mu=0}^{m-\lambda}
\frac{s_{[n-m-1]}l_{[\mu]}}{2^{m-\mu}}.
\end{equation}

We are now ready to apply the banded quantum Fourier transform to register $I$ of the initial 
state $|\psi_{i}\rangle$[see (\ref{Shor10})] and obtain with (\ref{UQFT_b}) and (\ref{B(s,l)})
\begin{align}
\hat{U}_{b}^{(QFT)}|\psi_i\rangle &=
\hat{U}_{b}^{(QFT)} \frac{1}{\sqrt{K}} \sum_{k=0}^{K-1} |s_{k}\rangle \cr
&= \frac{1}{\sqrt{K}} \sum_{k=0}^{K-1} \sum_{l=0}^{2^n-1} B(s_{k},l) |l\rangle \cr
&= \frac{1}{\sqrt{2^{n}K}} \sum_{k=0}^{K-1} \sum_{l=0}^{2^{n}-1}
\exp{\big\{ 2\pi i[S_{0}(s_{k},l) - S_{b+1}(s_k,l)] \big\}} |l\rangle.
\end{align}
From this we obtain
\begin{equation}
\label{rawpeakprob}
\tilde{P}_{j}(n,b,\omega) = \frac{1}{2^{n}K} \Bigg| \sum_{k=0}^{K-1} \exp{\big\{ 2\pi i
[S_{0}(s_{k},l_{j}) - S_{b+1}(s_{k},l_{j})] \big\}}\Bigg|^{2},
\end{equation}
which, using the expanded form (\ref{S(s,l)}) of $S$, can be written in the form
\begin{equation}
\label{RawProb}
\tilde{P}_{j}(n,b,\omega) = \frac{1}{2^{n}K} \Bigg|\sum_{k=0}^{K-1} 
e^{i[\Phi(n,s_{k},l_{j})-\varphi(n,b,s_{k},l_{j})]} \Bigg|^{2},
\end{equation}
where
\begin{equation}
\label{Phi}
\Phi(n,s,l) = \pi \sum_{m=0}^{n-1} \sum_{\mu=0}^{m} \frac{s_{[n-m-1]}l_{[\mu]}}{2^{m-\mu}}
\end{equation}
and
\begin{equation}
\label{phi}
\varphi(n,b,s,l) = \pi \sum_{m=b+1}^{n-1} \sum_{\mu=0}^{m-b-1} 
\frac{s_{[n-m-1]}l_{[\mu]}}{2^{m-\mu}}.
\end{equation}
While $\Phi$ in (\ref{Phi}) is already in a form useful for numerical 
calculations, we now derive an expression for $\exp{(i\Phi)}$,
which is more convenient for the analytical calculations in Sec.~\ref{SectionAR}. 
We start by summing (\ref{Phi}) in reverse order over $m$ ($n-m-1$ 
$\rightarrow$ $m$) to obtain:
\begin{align}
\label{Phi2}
\Phi(n,s,l) &= \pi \sum_{m=0}^{n-1} \sum_{\mu=0}^{n-m-1} 
\frac{s_{[m]}l_{[\mu]}}{2^{n-1-m}2^{-\mu}} \cr
&= \frac{\pi}{2^{n-1}} \sum_{m=0}^{n-1} 2^{m}s_{[m]} \sum_{\mu=0}^{n-m-1} 
2^{\mu}l_{[\mu]}.
\end{align}
If we extend the $\mu$ sum in (\ref{Phi2}) to include terms ranging from
$\mu = n-m$ to $\mu = n-1$, we notice that these extra terms generate 
even multiples of $2\pi$ in (\ref{Phi2}). Therefore, when computing 
$\exp{(i\Phi)}$, we can safely extend the $\mu$ sum to $\mu = n-1$,
since the extra terms, generating even multiples of $2\pi i$ in the argument
of the exponential function, do not contribute to $\exp{(i\Phi)}$. Therefore,
we obtain:
\begin{equation}
\exp{[i\Phi(n,s,l)]} = \exp{\bigg(\frac{\pi i}{2^{n-1}} \sum_{m=0}^{n-1} 2^{m}
s_{[m]} \sum_{\mu=0}^{n-1} 2^{\mu} l_{[\mu]}\bigg)}.
\end{equation}
Using the fact that
\begin{equation}
\sum_{m=0}^{n-1} 2^{m}s_{[m]} = s \mod 2^n,
\end{equation}
and similarly for $l$, we obtain
\begin{equation}
\exp{[i\Phi(n,s,l)]} = \exp{\bigg\{\frac{2\pi i}{2^n} \big[(s \bmod 2^n)(l \bmod 2^n)\big]\bigg\}}.
\end{equation}
The factor $2\pi i/2^n$ in the exponent induces a modulo operation and
we may also write
\begin{equation}
\label{ePhi}
\exp{[i\Phi(n,s,l)]} = \exp{\bigg\{\frac{2\pi i}{2^n} \big[(s \bmod 2^n)(l \bmod 2^n)
\big] \bmod 2^n\bigg\}}.
\end{equation}
Using the formula
\begin{equation}
[(A \bmod M)(B \bmod M)] \mod M = (A\cdot B)\mod M
\end{equation}
of elementary modular arithmetic, we may write (\ref{ePhi}) in the form:
\begin{equation}
\exp{[i\Phi(n,s,l)]} = \exp{\bigg[\frac{2\pi i}{2^n} (s\cdot l) \bmod 2^n\bigg]}.
\end{equation}
Now, we use (\ref{peakloc}) and (\ref{Shor9}) with $s_0 = 0$ to obtain:
\begin{equation}
\label{ePhi2}
\exp{[i\Phi(n,s_k,l_j)]} = \exp{\bigg[\frac{2\pi i}{2^n} (k2^{n}j + k\omega \beta_{j})
\bmod 2^n\bigg]}.
\end{equation}
The first term in parentheses contributes nothing to (\ref{ePhi2}), since it is
an integer and together with the prefactor in the exponent of (\ref{ePhi2}), 
amounts to an even multiple of $2\pi i$. Therefore, (\ref{ePhi2}) reduces to
\begin{equation}
\label{ePhi3}
\exp{[i\Phi(n,s_k,l_j)]} = \exp{\bigg[\frac{2\pi i}{2^n} (k\omega \beta_j) \bmod 2^n\bigg]}.
\end{equation}
Since $k\omega \leq 2^n$ and $|\beta_j| < \frac{1}{2}$, we have
$|k\omega \beta_j| < 2^n$. Therefore, the modulo operation in (\ref{ePhi3})
is not needed any more and we obtain
\begin{equation}
\exp{[i\Phi(n,s_k,l_j)]} = \exp{\bigg[2\pi i\bigg(\frac{k\omega \beta_j}{2^n}\bigg)\bigg]}.
\end{equation}
Thus we obtained a closed-form, analytical expression for $\exp(i\Phi)$.

Although [because of the presence of $\varphi(n,b,s_k,l_j)$ in (\ref{RawProb})] not
useful for the exact evaluation of (\ref{RawProb}), a well-justified approximation performed 
in Sec.~\ref{SectionAR} allows us to compute
\begin{equation}
\label{BigOmega}
\Omega(n,l_j,\omega) = \sum_{k=0}^{K-1} \exp{[i\Phi(n,s_k,l_j)]}
\end{equation}
separately.
Using the formula for computing geometric sums, we obtain:
\begin{align}
\Omega(n,l_j,\omega) &= \sum_{k=0}^{K-1} [\exp{(2\pi i\omega \beta_j/2^n)}]^k \cr
&= \frac{1-\exp{(2\pi i\omega \beta_{j}K/2^n)}}{1-\exp{(2\pi i\omega \beta_{j}/2^n)}}.
\end{align}
With (\ref{K}) we obtain
\begin{equation}
\Omega(n,l_j,\omega) \approx \frac{1-\exp{(2\pi i \beta_j)}}{1-\exp{(2\pi i \beta_{j} \omega/2^n)}}
\approx e^{i\pi \beta_j} K \frac{\sin{(\pi \beta_j)}}{(\pi \beta_j)}.
\end{equation}
Since $\varphi(n,b=n-1,s,l)=0$, we note in passing that
\begin{equation}
\label{P-Omega}
\tilde{P}_j (n,b=n-1,\omega) = \frac{1}{2^n K} |\Omega(n,l_j,\omega)|^2.
\end{equation}
We also need an analytical expression for the maximum value $\varphi_{max}(n,b)$ of
$\varphi(n,b,s_k,l_j)$, defined as
\begin{equation}
\label{phimaxdef}
\varphi_{max}(n,b) = \max_{k,j} \varphi(n,b,s_k,l_j).
\end{equation}
From (\ref{phi}) it is clear that $\varphi_{max}$ is obtained by setting
all $s_{[n-m-1]}$ and $l_{[\mu]}$ values equal to $1$. This procedure yields
\begin{equation}
\label{phimaxsum}
\varphi_{max}(n,b) = \pi \sum_{m=b+1}^{n-1} \sum_{\mu=0}^{m-b-1} \frac{1}{2^{m-\mu}}.
\end{equation}
Only the formula for evaluating geometric sums is needed to compute the
value of $\varphi_{max}$ in (\ref{phimaxsum}). We obtain
\begin{equation}
\label{phimax}
\varphi_{max}(n,b) = 2\pi [2^{-b-1}(n-b) - 2^{-b} + 2^{-n}].
\end{equation}
We now show that a quantum computer 
performs perfectly, no matter what $b$ is, if $\omega$ is 
a power of 2, i.e., 
\begin{equation}
P(n,b,\omega) = 1,\ \ \ {\rm for}\ \omega=2^{\alpha}, \ \ \alpha\geq 0\ {\rm integer}. 
\label{PERFM99}
\end{equation}
For such an $\omega$, we notice that (i) the $\kappa$th binary digit
of any $l_j$ is zero for $\kappa \leq n-\alpha$ since according to (\ref{peakloc})
\begin{equation}
l_j = 2^{n-\alpha} j, \ \ \ j = 0,1, \dots , \omega-1
\end{equation}
is already integer, which implies $\beta_j = 0$, and (ii) the $\iota$th
binary digit of any equivalence class element in $[s_0]$ [see (\ref{Shor9})] for
$0 \leq \iota < \alpha$ is identical to that of $s_0$. Thus, we write $\varphi(n,b,s,l)$ in
(\ref{phi}) in the form
\begin{align}
\varphi(n,b,s,l) &= \pi\Bigg(\sum_{m=n-\alpha+b+1}^{n-1} \sum_{\mu=0}^{m-b-1}
\frac{s_{[n-m-1]}l_{[\mu]}}{2^{m-\mu}} + \sum_{m=b+1}^{n-\alpha+b} \sum_{\mu=0}
^{m-b-1} \frac{s_{[n-m-1]}l_{[\mu]}}{2^{m-\mu}}\Bigg) \cr
&=\begin{cases} 0, &\mbox{if } \alpha \leq b+1, \cr
\pi \sum_{m=n-\alpha+b+1}^{n-1} \sum_{\mu=n-\alpha}^{m-b-1}
\frac{s_{[n-m-1]}l_{[\mu]}}{2^{m-\mu}}, & \mbox{if } \alpha > b+1, \end{cases}
\end{align}
where the second equality was obtained by using (i). Now, we observe that
the $n-m-1$th digit of $s$ is bounded between $0$ and $\alpha-b-2$ inclusively.
Then, using (ii), we obtain
\begin{align}
\varphi(n,b,s=s_k,l=l_j) &= \pi \sum_{m=n-\alpha+b+1}^{n-1} \sum_{\mu=n-\alpha}^{m-b-1}
\frac{(s_k)_{[n-m-1]}(l_j)_{[\mu]}}{2^{m-\mu}} \cr
&= \pi \sum_{m=n-\alpha+b+1}^{n-1} \sum_{\mu=n-\alpha}^{m-b-1}
\frac{(s_0)_{[n-m-1]}(l_j)_{[\mu]}}{2^{m-\mu}} \cr &= \tilde{\varphi}_j,
\label{idealpeakphi}
\end{align}
where $\tilde{\varphi}_j$ is a constant for any $s_k$ and a given $l_j$. Inserting 
(\ref{idealpeakphi}) in (\ref{RawProb}), $\tilde{P}_j (n,b,\omega)$ becomes
\begin{align}
\tilde{P}_j (n,b,\omega) 
&= \frac{1}{2^n K} \Bigg| \sum_{k=0}^{K-1} e^{i[\Phi(n,s_k,l_j)-\tilde{\varphi}_j]} 
\Bigg|^2  \cr &=  \frac{1}{2^n K} \big| e^{-i\tilde{\varphi}_j}\big|^2 \Bigg| \sum_{k=0}^{K-1} e^{i\Phi(n,s_k,l_j)}\Bigg|^2 \cr &= \frac{1}{2^n K} \Bigg| \Omega(n,l_j,\omega) \Bigg|^2
= \tilde{P}_j (n,b=n-1,\omega),
\label{idealpeakprob}
\end{align}
where we used (\ref{BigOmega}) and (\ref{P-Omega}).
With (\ref{rawmeasure}) and (\ref{idealpeakprob}) we obtain
\begin{equation}
\tilde{P}(n,b,\omega) = \sum_{j=0}^{\omega-1} \tilde{P}_j (n,b=n-1,\omega) 
= \tilde{P}(n,b=n-1,\omega).
\end{equation}
Therefore, with (\ref{measure}), the normalized probability (the performance measure)
$P(n,b,\omega)$ reads
\begin{equation}
P(n,b,\omega) = \frac{\tilde{P}(n,b=n-1,\omega)}{\tilde{P}(n,b=n-1,\omega)} = 1,
\end{equation}
which completes the proof.

Since $\omega=2$ always exists (see Appendix~\ref{AppendixA}), 
this is an important observation, since the corresponding quantum computer
works perfectly in this case for any $n$ and any $b$. The trick, of course, 
is to find the seed $x$ that yields $x^2\mod N=1$. 
This, however, is an unsolved problem for large $N$. 

\begin{figure}
\centering
\includegraphics[scale=1]{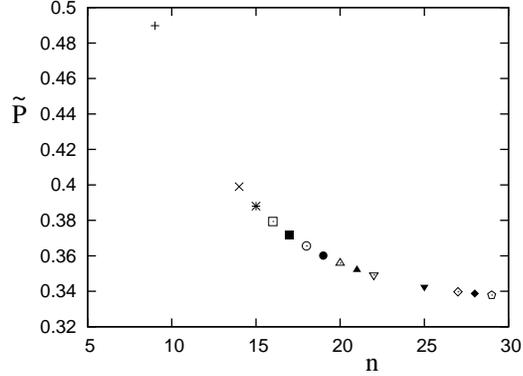}
\caption{\label{order6} 
Probability $\tilde{P}(n,b=1,\omega=6)$ as a function of $n$
for 14 different semiprimes $N$ with seeds chosen such that
$\omega=6$. As expected, the data clearly asymptotes to the value $1/3$.
  }
\end{figure}

If $\omega$ is not a power of 2, we write it in the form
\begin{equation}
\label{orderdecomp}
\omega = r 2^\alpha,\,\,\,\,\,\, r,\,\alpha \ {\rm integer},
\end{equation}
where $r$ is odd. For such an $\omega$, according to (\ref{peakloc}), we
may write $l_j$ as
\begin{equation}
l_j = \bigg(\frac{2^{n-\alpha}}{r}\bigg) j + \beta_j.
\end{equation}
Therefore, if $j$ is a multiple of $r$, we have $\beta_j = 0$ and
$\tilde{P}_j (n,b,\omega) = 1/\omega$, which is proved by following
the corresponding steps for the case where $\omega$ is a power of 2.
This means that the contribution
of these $j$ values to $\tilde{P} (n,b,\omega)$ is $1/r$. 
This is a constant contribution, which does not depend on
either $n$ or $b$. Therefore, if for large $n$ the contributions to $\tilde{P}(n,b,\omega)$
tend to zero for the $l_j$ peaks for which $j$ is not
a multiple of $r$, we expect $\tilde{P}(n,b,\omega)$ to approach $1/r$ 
for large $n$.
This is demonstrated in Fig.~\ref{order6}, which shows $\tilde{P}(n,b=1,\omega=6)$
as a function of $n$. Since in this case $\omega=3\times2^1$,
we expect $\tilde{P}(n,b=1,\omega=6)$ to approach $1/3$, which is clearly
confirmed in Fig.~\ref{order6}.
%
\section{Numerical Results} 
\label{SectionNR}

%
\begin{figure}
\centering
\includegraphics[scale=1]{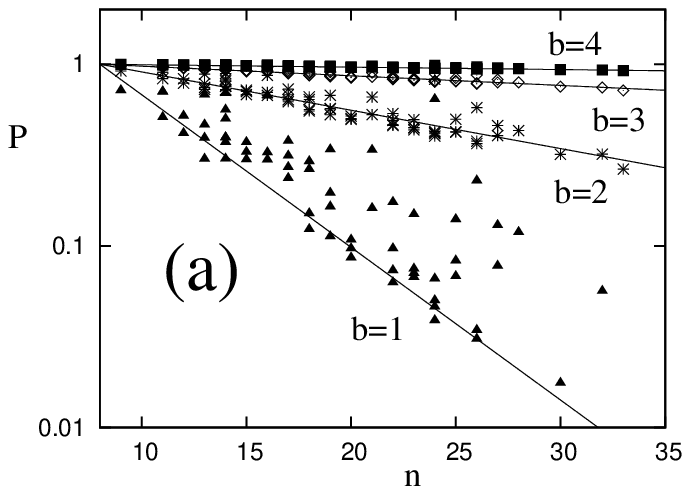} \\
\includegraphics[scale=1]{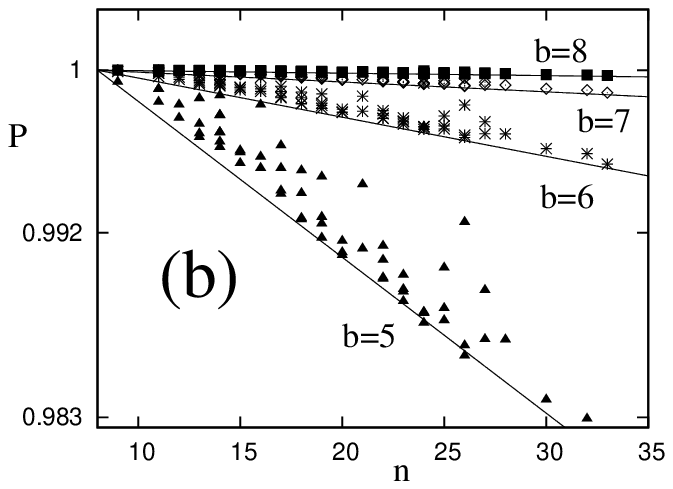}
\caption{\label{fig2} 
Normalized probability $P$, represented by the
properly averaged performance measure (\ref{NUMRES2}),
for successful factorization of sample semiprimes $N$ of binary length
$\log_2{(N)} \sim n/2$ as a function of $n$ for several
bandwidths $b$, ranging from $b=1$ to $b=8$. 
(a) $b=1$ (triangles),
$b=2$ (stars), $b=3$ (diamonds), and $b=4$ (squares).
(b) $b=5$ (triangles),
$b=6$ (stars), $b=7$ (diamonds), and $b=8$ (squares).
The solid lines through the data points are the fit 
functions (\ref{expfit}).
Notice the visual similarity of (a) and (b), which illustrates
the exponential scaling of $\xi_b$ in $b$.
  }
\end{figure}
%
\begin{figure}
\centering
\includegraphics[scale=1]{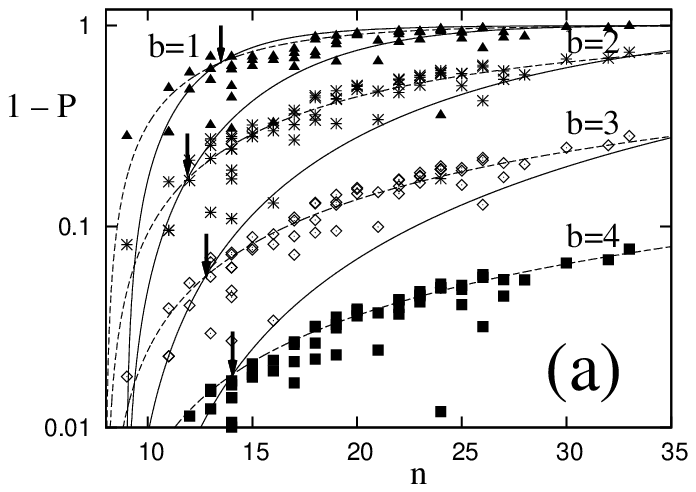} \\
\includegraphics[scale=1]{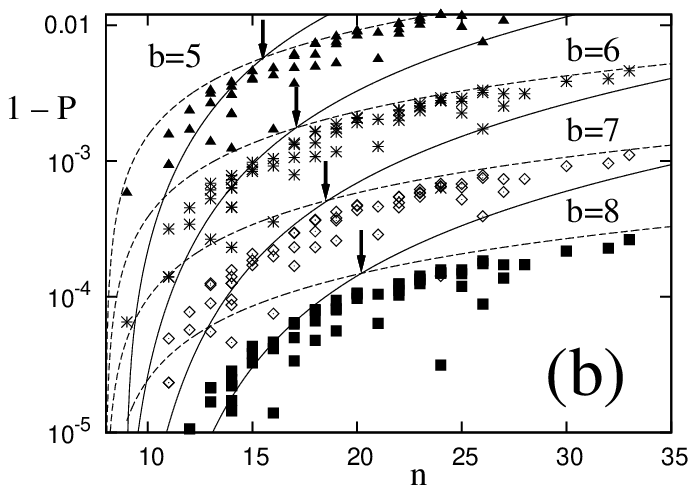}
\caption{\label{fig3}    Small-$n$ behavior of
$1-P$ [see (\ref{NUMRES2})]
for several sample semiprimes $N$ (plot symbols)
with proper average over 
$\{\omega(N)\}$. The bandwidth $b$ ranges from $b=1$
to $b=8$.
(a) $b=1$ (triangles),
$b=2$ (stars), $b=3$ (diamonds), and $b=4$ (squares).
(b) $b=5$ (triangles),
$b=6$ (stars), $b=7$ (diamonds), and $b=8$ (squares).
The solid lines are the non-exponential fit functions
(\ref{gaussfit}).
The dashed lines are the fit functions (\ref{expfit}). 
The cross-over points between the small-$n$, non-exponential
behavior and the 
large-$n$, exponential behavior
[i.e. the intersections of (\ref{expfit}) and (\ref{gaussfit})]
are marked by arrows. 
       }
\end{figure}
In this section we explore, numerically, the performance of 
Shor's algorithm supplied with a banded quantum Fourier transform 
of bandwidth $b$. The performance is measured objectively 
with the help of the quantitative performance measure 
$P(n,b,\omega)$ defined in (\ref{measure}). 
In contrast to a similar investigation 
by Fowler and Hollenberg \cite{FH},
who use an effective $\omega$ for the investigation of the 
performance of the banded Shor algorithm, we opted for 
a more realistic simulation of the performance of Shor's algorithm 
using ensembles of semiprimes $N$ together with their 
exact associated orders $\omega$. Thus, our procedure 
for computing the performance measure is as follows. For given $n$ 
we choose an ensemble of semiprimes $N=pq$ such that 
\begin{equation}
n = \lfloor 2\log_2(N) + 1 \rfloor , 
\label{NUMRES1}
\end{equation}
where $\lfloor\ldots\rfloor$ is the floor function \cite{floor}. 
This ensures that $n$ is at least twice as large as the
number of binary digits of $N$, as required to reliably
determine the order $\omega$ with an $n$-qubit 
quantum computer \cite{ShorSIAM, Mermin, EkJo}.
For each $N$ we compute its set of orders 
$\{\omega_1,\ldots,\omega_{a(N)}\}$, where $a(N)$ is 
the number of orders for given $N$. We also define 
the multiplicity of a given order $\omega$ as the 
number $\nu(\omega)$ of seeds $x$ of order $\omega$. 
Thus equipped, we compute the performance 
$P_N(n,b)$ as the properly weighted average 
\begin{equation}
    P_N(n,b) = \frac{1}{\varphi_E(N)} \sum_{j=1}^{a(N)} 
                  \nu(\omega_j) P(n,b,\omega_j),
\label{NUMRES2}
\end{equation}
where $P(n,b,\omega)$ is defied in (\ref{measure}) and
$\varphi_E(N)$ is Euler's totient function \cite{Jacobson}. 
  
In Fig.~\ref{fig2}~(a) we show $P_N(n,b)$ for various choices of $N$
for $b=1,\ldots,4$ and $n$ 
ranging from $n=9$ to $n=33$. Plot symbols correspond 
to particular $N$ values and there are up to 7 
semiprimes $N$ per $n$. Overall we see that the data exhibit 
exponential behavior on average, which is well represented by the fit 
lines 
\begin{equation}
\label{expfit}
P_{>}(n,b) = 2^{-\xi_b(n-8)},\ \ \ \xi_b = 1.1\times 2^{-2b}
\end{equation}
drawn through the data points.
In Sec.~\ref{SectionARB} we present an analytical model that explains
the $b$-scaling of (\ref{expfit}) and in addition reproduces the pre-factor in (\ref{expfit})
within $10\%$.
Figure~\ref{fig2}~(b) shows corresponding data 
for $b=5,\ldots,8$. Again, the data points behave exponentially 
and are well approximated by the fit lines defined in (\ref{expfit}). 
This illustrates that the $b$ and $n$ scaling in (\ref{expfit}) holds over a considerable
range of $b$ and $n$ values.
 
While on the large scale of Fig.~\ref{fig2} the data show an 
exponential behavior, looking more closely at the small-$n$ 
regime, we see definite deviations from exponential 
behavior. Plotting $1-P(n,b)$ magnifies the $P(n,b)$ 
behavior in the small-$n$ region and clearly brings out the 
deviations from exponential behavior. This is illustrated 
in Fig.~\ref{fig3}, which shows the data of Fig.~\ref{fig2},
plotted as $1-P(n,b)$. The dashed lines in 
Fig.~\ref{fig3} are the exponential fit lines defined in 
(\ref{expfit}). We see that even on this 
magnified scale and in the large-$n$ regime the data are well represented
by the exponentials (\ref{expfit}). For small $n$, 
however, the data clearly deviate from exponential, but 
are well fit by the solid lines representing the function \cite{Nam}
\begin{equation}
\label{gaussfit}
P_{<}(n,b)=\tilde{P}_{<}(n,b) / \bar{f},
\end{equation}
where
\begin{equation}
\bar{f} = \int_{-1/2}^{1/2} \frac{\sin^2 (\pi \beta)}{(\pi \beta)^2} d\beta \approx 0.774
\end{equation}
and
\begin{equation}
\label{rawproblittle}
\tilde{P}_{<} (n,b) = \langle\frac{1}{r} \rangle + \Big( 1- \langle \frac{1}{r} \rangle \Big)
\bigg(\frac{\bar{f} - \langle
\frac{1}{r} \rangle}
{1- \langle \frac{1}{r} \rangle} \bigg) \exp{[-\varphi^2_{max} (n,b)/100]},
\end{equation}
where $\varphi_{max}$ is given in (\ref{phimax}), $r$ is defined in (\ref{orderdecomp}),
and $\langle \frac{1}{r} \rangle = 2^{-(n-8)/2.6}$ (see Appendix~ \ref{AppendixC}).
Based on our numerical evidence, we 
conclude that $P(n,b)$ shows a clear transition from 
non-exponential behavior for small $n$ to exponential 
behavior for large $n$. The arrows in Fig.~\ref{fig3}
point to the locations of the transition between the 
two regimes and are the intersection points between 
the functions defined in (\ref{expfit}) and (\ref{gaussfit}).

Combining expressions (\ref{expfit}) and (\ref{gaussfit}),
we derive an analytical expression, $n_t(b)$,
for the transition points between the two different regimes
for given $b$.
The transition points $n_t$ are defined as the $n$-value at which
(\ref{expfit}) equals (\ref{gaussfit}). A useful analytical formula,
approximately valid for $b\gtrsim8$, is obtained in the following way.
For $b\gtrsim8$, we noticed numerically that the $1/r$ terms in (\ref{rawproblittle})
may be neglected, resulting only in a small shift of $n_t$ of about 2 units
in $n$. Therefore, to lowest order, $P_{<}(n_t,b) = P_{>}(n_t,b)$ results in
\begin{equation}
\frac{\varphi^2_{max}(n_t,b)}{100} = \xi_b \ln{(2)} (n_t-8),
\end{equation}
which implies
\begin{equation}
\label{crossrelation}
1.1 \times 2^{-2b} \ln{(2)} (n_t-8) = \frac{4\pi^2}{100}
\big[ 2^{-b-1} (n_t-b-2) + 2^{-n_t} \big]^2.
\end{equation}
At this point we notice that the transitions $n_t$ between the two regimes
occur at $n$ values for which
\begin{equation}
\label{crosscondition}
2^{-n_t} \ll 2^{-b},
\end{equation}
which implies that we can safely neglect the $2^{-n_t}$ term in (\ref{crossrelation}).
This turns (\ref{crossrelation}) into the quadratic equation
\begin{equation}
\label{crossquadratic}
n_t^2 - 2n_t(C+b+2) + 16C + (b+2)^2 = 0,
\end{equation}
where we defined
\begin{equation}
C = \frac{55 \ln{(2)}}{\pi^2}.
\end{equation}
Solving (\ref{crossquadratic}) yields
\begin{equation}
\label{crossformula}
n_t = b + 5.9 + \sqrt{7.7(b+2) - 47}.
\end{equation}
%
\begin{figure}
\centering
\includegraphics[scale=1]{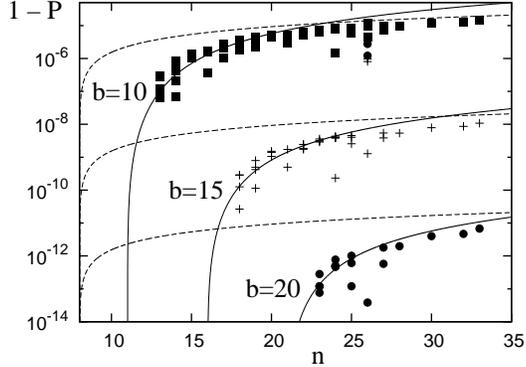}
\caption{\label{fig4}    Small-$n$ behavior of semiprimes $N$ 
for $b=10$ (squares), $b=15$ (crosses), and $b=20$ (bullets). 
The full lines are the non-exponential performance functions 
$P_{<}(n,b)$ [see (\ref{gaussfit})].
The dashed lines are the corresponding large-$n$, 
exponential fit functions (\ref{expfit}). 
       }
\end{figure}
The expression (\ref{crossformula})
for the transition points shows that the onset of exponential
behavior is shifted toward larger $n$ for larger $b$.
Formula (\ref{crossformula}) for the transition points $n_t(b)$ is 
useful for extrapolating into the 
practically relevant 
qubit regime $n\gtrsim 4000$, where classical 
computers cannot follow any more. In this 
classically inaccessible regime, we can then 
decide on the basis of (\ref{crossformula}),
e.g., whether for given $b$ and 
very large $n$, formula (\ref{expfit}) 
or formula (\ref{gaussfit}) should 
be used to predict the performance of the 
quantum computer. 
For $b = 1, \ldots, 4$, as shown in Fig.~\ref{fig3}~(a), the transition
is poorly defined, whereas, as shown in Fig.~\ref{fig3}~(b), the
transition is progressively
better defined as $b$ increases. That this trend continues is shown
in Fig.~\ref{fig4}, which shows data for $b=10,\,15$, and 20.
We also see that the quality of the fit of the data with (\ref{gaussfit})
improves for increasing $b$. The sharp cut-off displayed by $P_{<}(n,b)$
in Fig.~\ref{fig4} at $n=11$ $(b=10)$, $n=16$ $(b=15)$, and
$n=22$ $(b=20)$ is also understood since, according to (\ref{phimax}),
$\varphi_{max}(n,b) = 0$ for $n=b+1$.


\section{Analytical Results}
\label{SectionAR}

Our analytical investigation of the performance measure starts with (\ref{RawProb}). Analytically
and numerically we found that $\Phi(n,s_k,l_j)$
is a slow function of $k$, whereas
$\varphi(n,b,s_k,l_j)$ is a fast, erratic function of $k$. Therefore, we can write
approximately, 
\begin{align}
\label{separabilitymodel}
	\tilde{P}_{j}(n,b,\omega) &\approx \frac{1}{2^n K} \Bigg| 
	\bigg[ \sum_{k=0}^{K-1} e^{i\Phi(n,s_k,l_j)} \bigg]
	\langle e^{-i\varphi} \rangle_{n,b,l_j} \Bigg|^2 \cr
	&= \frac{1}{2^n K} \Big|\Omega(n,l_j,\omega)
	\Big|^2 \Big|\langle e^{-i\varphi} \rangle_{n,b,l_j}\Big|^2,
\end{align}
where $\Omega(n,l_j,\omega)$ is defined in (\ref{BigOmega}) and
\begin{equation}
\label{avgemip}
\langle e^{-i\varphi} \rangle_{n,b,l_j} = \frac{1}{K} \sum_{k=0}^{K-1} e^{-i\varphi(n,b,s_k,l_j)}.
\end{equation}
With (\ref{measure}), (\ref{rawmeasure}), and (\ref{P-Omega}) we now obtain
\begin{equation}
\label{k-sepresult}
P(n,b,\omega) = \frac{\displaystyle \sum_{j=0}^{\omega-1} \Big|\Omega(n,l_j,\omega)
	\Big|^2 \Big|\langle 
	e^{-i\varphi} \rangle_{n,b,l_j}\Big|^2}
	{\displaystyle \sum_{j=0}^{\omega-1} \Big|\Omega(n,l_j,\omega)\Big|^2}.
\end{equation}
We now proceed with a slightly less but still extremely accurate approximation
by separating (\ref{k-sepresult}) in $j$, which then yields
\begin{equation}
\label{analyticalbasis}
P(n,b,\omega) = \frac{1}{\omega} \sum_{j=0}^{\omega-1} \big|\langle e^{-i\varphi} \rangle_{n,b,l_j}
\big|^2 = \langle  \big|\langle e^{-i\varphi} \rangle_k \big|^2  \rangle_{j},
\end{equation}
where $\langle \ldots \rangle_k$ and $\langle \ldots \rangle_j$ are averages over $k$ and $j$,
respectively.
This expression for the performance measure $P(n,b,\omega)$ is the basis
of our analytical work.

Since (\ref{analyticalbasis}) is based on the validity of the separation in $k$ and $j$,
both are investigated in detail in Sec.~\ref{SectionARA}. A random model is used
in Sec.~\ref{SectionARB} to evaluate (\ref{analyticalbasis}) analytically in the large-$n$
regime. This yields an analytical explanation for the $b$-scaling in (\ref{expfit}) and excellent
agreement with the prefactor of the exponential term in (\ref{expfit}). In Sec.~\ref{SectionARC},
again assuming separation in $k$ and $j$, we then arrive at an analytical formula describing 
the small-$n$ regime, which predicts the functional form and the $b$-scaling
of (\ref{gaussfit}) very well, and also provides an estimate of the overall scaling factor.

\subsection{Separability}
\label{SectionARA}

In this section we investigate in detail the quality of the separations in $k$ and in $j$,
which lead to our jump-off point (\ref{analyticalbasis}) for the analytical calculations
reported in Sec.~\ref{SectionARB} and Sec.~\ref{SectionARC}.
\begin{figure}
\centering
\includegraphics[scale=1]{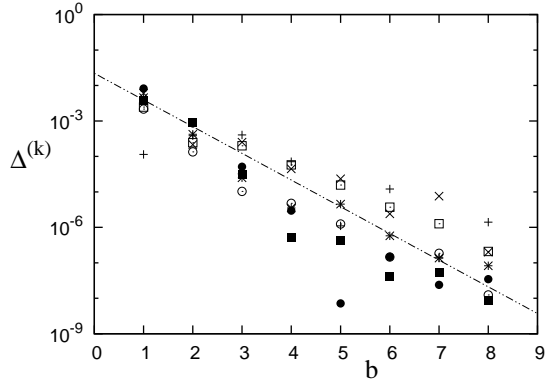}
\caption{\label{fig5}
	Relative error $\Delta^{(k)}$ of $k$ separation
	as a function of $b$ for several semiprimes $N$.
	The data shows that the error is negligible. The
	fit line $\Delta = 2^{-2.5b-5.5}$ (dashed line) shows
	that the relative error vanishes exponentially in $b$.
       }
\end{figure}

We start with justifying the separation in $k$. To this end we define
\begin{equation}
A^{(k)} = \sum_{j=0}^{\omega-1} \Bigg| \sum_{k=0}^{K-1} e^{i\Phi(n,s_k,l_j) - 
	i\varphi(n,b,s_k,l_j)} \Bigg|^2
\end{equation}
and
\begin{align}
B^{(k)} &= \sum_{j=0}^{\omega-1} \Bigg| \Bigg[\sum_{k=0}^{K-1} e^{i\Phi(n,s_k,l_j)} \Bigg] 
	\frac{1}{K} \sum_{k'=0}^{K-1} e^{-i\varphi(n,b,s_{k'},l_j)} \Bigg|^2 \cr
	&= \sum_{j=0}^{\omega-1} \Big|\Omega(n,l_j,\omega)\Big|^2 
	\Big|\langle e^{-i\varphi} \rangle_{n,b,l_j} \Big|^2
\end{align}
and compute the relative error
\begin{equation}
\Delta^{(k)} = \frac{\big|A^{(k)} - B^{(k)} \big|}{\big| A^{(k)} \big|}
\end{equation}
incurred by the $k$ separation.
Figure~\ref{fig5} shows $\Delta^{(k)}$ as a function of $b$ for various choices of $N$. 
We clearly see that $k$ separation is an excellent approximation,
which produces negligible, exponentially small errors. We plotted the line
$\Delta = 2^{-2.5b-5.5}$ through the data to guide the eye. This line shows that the relative error
of $k$ separation vanishes exponentially in $b$.
\begin{figure}
\centering
\includegraphics[scale=1]{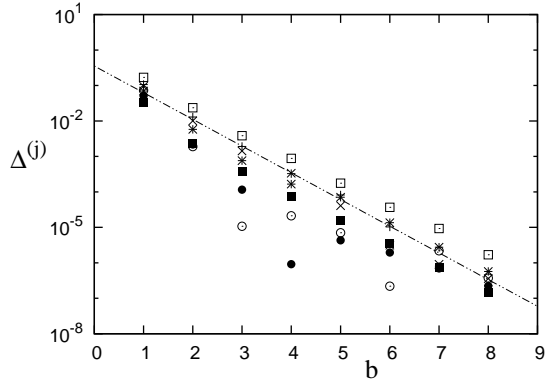}
\caption{\label{fig6}
	Relative error $\Delta^{(j)}$ of $j$ separation
	as a function of $b$ for several semiprimes $N$.
	A fit line, $\Delta = 2^{-2.5b-1.5}$ (dashed line)
	is also shown.
	Compared with $k$ separation (see Fig.~\ref{fig5})
	the error decays with the same exponent, only the
	overall scale factor is different.
       }
\end{figure}

Turning now to the $j$ separation, we define
\begin{equation}
A^{(j)} = B^{(k)}
\end{equation}
and
\begin{equation}
B^{(j)} = \Bigg[\sum_{j=0}^{\omega-1} \Big| \Omega(n,l_j,\omega) \Big|^2 \Bigg] \frac{1}{\omega}
	      \sum_{j=0}^{\omega-1} \Big| \langle e^{-i\varphi} \rangle_{n,b,l_j} \Big|^2
\end{equation}
and compute the relative error of $j$ separation
\begin{equation}
\Delta^{(j)} = \frac{\big|A^{(j)} - B^{(j)}\big|}{\big| A^{(j)} \big|}.
\end{equation}
Figure~\ref{fig6} shows $\Delta^{(j)}$ as a function of $b$ for various choices of $N$.
Apparently, while a bit less accurate than $k$ separation, 
$j$ separation is still highly accurate, improving exponentially with $b$.
This is seen from the fit line $\Delta = 2^{-2.5b-1.5}$ through the data
in Fig.~\ref{fig6}, which also shows that $\Delta^{(k)}$ and $\Delta^{(j)}$
decay with the same exponential factor in $b$, and are offset by a 
constant only.

\subsection{Large-$n$, exponential regime}
\label{SectionARB}

In this section we evaluate (\ref{analyticalbasis}) analytically in a model
in which we treat $s_k$ and $l_j$ as independent random variables.
This model, obviously, cannot capture the correlations between $s_k$
and $l_j$ introduced by $\omega$ and yields $P(n,b,\omega)$ that is 
independent of $\omega$. Therefore, the $\omega$-average in (\ref{NUMRES2})
is trivial and $P_N (n,b)$ does not depend on $N$ either. Therefore,
we write $P_N (n,b) \rightarrow P(n,b)$ as the prediction of the random model.
However, even in this model, where $\omega$-correlations are entirely neglected,
it is hard to evaluate the expectation value of the
exponential. Therefore, we proceed to evaluate (\ref{analyticalbasis}) via its moment expansion
\begin{equation}
\label{momentexpansion}
\langle \big| \langle e^{-i\varphi} \rangle_k \big|^2 \rangle_j = 
1 - \big[ \langle \varphi^2 \rangle_{kj} - \langle \langle \varphi \rangle_{k}^{2}\rangle_j \big]
+ \bigg[ \frac{1}{12} \langle \varphi^4 \rangle_{kj} + \frac{1}{4} \langle\langle \varphi^2 \rangle_k^2
\rangle_j - \frac{1}{3} \langle \langle \varphi \rangle_k \langle \varphi^3 \rangle_k
\rangle_j \bigg] \pm \dots,
\end{equation}
where we used $\langle \ldots \rangle_{kj} = \langle\langle \ldots \rangle_k \rangle_j =
\langle \langle \ldots \rangle_j \rangle_k$
in cases where the averages commute.
We start by computing
\begin{equation}
\label{sqphiavgsum}
\langle \varphi^2 \rangle_{kj} = \pi^2 \sum_{m,m' = b+1}^{n-1} \sum_{\mu=0}^{m-b-1}
	\sum_{\mu'=0}^{m'-b-1} \frac{\langle s_{[n-m-1]}s_{[n-m'-1]} \rangle_k 
	\langle l_{[\mu]}l_{[\mu']} \rangle_j}	{2^{m+m'-\mu-\mu'}},
\end{equation}
where we made use of the assumed independence of $s$ and $l$. Taking
into account that the binary digits of $s$ and $l$ can only take the values
$0$ and $1$, we obtain
\begin{equation}
\label{doublesavg}
\langle s_{[\alpha]} s_{[\beta]} \rangle_k = \frac{1}{2} \delta_{\alpha\beta} 
+ \frac{1}{4}(1-\delta_{\alpha\beta}),
\end{equation}
and a similar expression for $\langle l_{[\mu]} l_{[\mu']}\rangle_j$. Because of (\ref{doublesavg}),
the evaluation of the quadruple sum (\ref{sqphiavgsum}) is lengthy, but can be performed
analytically. The result is
\begin{equation}
\label{sqphiavg}
\langle \varphi^2 \rangle_{kj} = \bigg( \frac{\pi^2}{144} \bigg) 2^{-2b} \Big[
9x^2 + 21x - 10 + 9(2+x)2^{-x} + 2^{-2x} \Big],
\end{equation}
where
\begin{equation}
\label{xdef}
x = n-b-2.
\end{equation}
Next, we evaluate $\langle \langle \varphi \rangle_k^2 \rangle_j$. With (\ref{doublesavg}) and
following the same procedures that lead to (\ref{sqphiavg}), we obtain
\begin{equation}
\label{sqlinphiavg}
\langle \langle \varphi \rangle_k^2 \rangle_j = \bigg( \frac{\pi^2}{96} \bigg) 2^{-2b} \Big[
6x^2 + 6x - 4 + 6(1+x)2^{-x} + 2^{-2x} \big],
\end{equation}
where $x$ is defined in (\ref{xdef}). We define
\begin{equation}
\label{vardef}
\hat{\sigma}^2 = \langle \varphi^2 \rangle_{kj} - \langle \langle \varphi \rangle_k^2 \rangle_j,
\end{equation}
which, on the basis of the results (\ref{sqphiavg}) and (\ref{sqlinphiavg}), is explicitly
given by
\begin{equation}
\label{variance}
\hat{\sigma}^2 = \bigg( \frac{\pi^2}{288} \bigg) 2^{-2b} \Big(24x - 8 
+ 18\times 2^{-x} - 2^{-2x} \Big).
\end{equation}
With (\ref{analyticalbasis}) and up to second order in the moment expansion 
(\ref{momentexpansion}), the performance measure is now given by
\begin{equation}
\label{varprob}
P(n,b) \approx 1-\hat{\sigma}^2.
\end{equation}
Comparing (\ref{varprob}) with the fit function (\ref{expfit}) and using (\ref{xdef}),
we see that (\ref{varprob}), to
leading order in $n$, is the first-order expansion of
\begin{equation}
\label{expanalytical}
P^{(a)} (n,b) \sim 2^{-\xi_{b}^{(a)} n},
\end{equation}
where
\begin{equation}
\xi_{b}^{(a)} = \bigg[ \frac{\pi^2}{12 \ln{(2)}} \bigg] \times 2^{-2b} \approx 1.19 \times 2^{-2b}.
\end{equation}
This analytical result recovers the $2^{-2b}$ scaling of the fit line (\ref{expfit}), and is within $10\%$
of the exponential prefactor in (\ref{expfit}).

The analytical evaluation of the 4th order terms in (\ref{momentexpansion})
is technically straightforward, but tedious,
and not essential at this point. Our numerical calculations show that the 4th order terms 
are approximately given by $(\hat{\sigma}^2)^2/2$, and
are therefore very small. This has two consequences: (i) it shows that up to 4th order in $\varphi$
the probability measure $P(n,b)$ for fixed $b$ is consistent with exponential decay in $n$ 
and (ii) that because of 
their smallness it is currently not necessary to evaluate the 4th order terms analytically.

To conclude this section, we compute
\begin{equation}
\langle \varphi \rangle_{kj} = \frac{\pi}{4} \sum_{m=b+1}^{n-1} \sum_{\mu=0}^{m-b-1} 
\frac{1}{2^{m-\mu}},
\end{equation}
which is needed in the following section.
Using the summation formula for the evaluation of geometric sums, we obtain
\begin{equation}
\label{linphiavg}
\langle \varphi \rangle_{kj} = \frac{\pi}{4} [2^{-b}(n-b-2) + 2^{1-n}] = \frac{1}{4} \varphi_{max},
\end{equation}
where we related $\langle \varphi \rangle_{kj}$ to $\varphi_{max}$ via (\ref{phimax}).

\subsection{Small-$n$, non-exponential regime}
\label{SectionARC}

Our starting point is again equation (\ref{analyticalbasis}), but in this section we focus on
the small-$n$ regime, i.e. $n < n_t(b)$ [see (\ref{crossformula})]. We first derive some useful
relations that can then be used to evaluate (\ref{analyticalbasis}) approximately in this regime.
We start by inspecting $\varphi(n,b,s,l)$ in (\ref{phi}). We notice that
\begin{equation}
\label{redefphi}
\varphi(n,b,s,l) = \frac{\pi}{2^{n-1}} \sum_{i=0}^{n-b-2} \big[ (2^i s_{[i]} l) \mod 2^{n-b-1} \big].
\end{equation}
Since the modulus of the product of two numbers is smaller than or equal to
the product of the moduli of two numbers, we obtain
\begin{align}
\label{sepphi}
\varphi(n,b,s,l) & \leq \frac{\pi}{2^{n-1}} \sum_{i=0}^{n-b-2} \big[ (2^i s_{[i]} \bmod 2^{n-b-1})
	(l \bmod 2^{n-b-1}) \big] \cr
	&= \frac{\pi}{2^{n-1}} \big[ (s \bmod 2^{n-b-1})(l \bmod 2^{n-b-1}) \big],
\end{align}
where the equality is obtained by using
\begin{equation}
\Bigg( \sum_{i=0}^{n-b-2} 2^i s_{[i]} \Bigg) \bmod 2^{n-b-1} 
= (s \bmod 2^{n-b-1})\bmod 2^{n-b-1} = s \bmod 2^{n-b-1} .
\end{equation}
In order to compensate for the difference between (\ref{redefphi}) and (\ref{sepphi}),
we introduce an effective parameter $\bar{l}$ in (\ref{sepphi}) such that
\begin{equation}
\label{lbardef}
\varphi = \frac{\pi}{2^{n-1}} (s \bmod 2^{n-b-1}) \bar{l} \leq \varphi_{max},
\end{equation}
where the inequality is obtained from the definition of $\varphi_{max}$ in (\ref{phimaxdef}).
Since this inequality must hold for any $s$, the inequality (\ref{lbardef}) implies
\begin{equation}
\label{lbarabscon}
\pi2^{-b}\bar{l} < \varphi_{max},
\end{equation}
where we used $\max (s \bmod 2^{n-b-1}) \approx 2^{n-b-1}$. Assuming
the random model used in Sec.~\ref{SectionARB}, in particular its assumption
of statistical independence of $s$ and $l$, we compute the average of (\ref{lbardef}).
With (\ref{linphiavg}) we obtain
\begin{equation}
\langle \varphi \rangle_{kj} = \frac{\varphi_{max}}{4} = \frac{\pi}{2^{n-1}} \langle s \bmod 2^{n-b-1}
\rangle_k \langle \bar{l} \rangle_j = \frac{\pi}{2} 2^{-b} \langle \bar{l} \rangle_j.
\end{equation}
Hence, solving for $\langle \bar{l} \rangle_j$, dropping the small term $2^{-n}$ in (\ref{phimax}), we expect
\begin{equation}
\label{avglbar}
\langle \bar{l} \rangle_j \simeq \frac{n-b-2}{2}.
\end{equation}
We note that $\langle \bar{l} \rangle_j$ in (\ref{avglbar}) fulfills (\ref{lbarabscon}).
Next, by writing the order of a seed as $\omega = 2^\alpha r$ [see (\ref{orderdecomp})],
and by using the form
of an element $s_k$ of an equivalence class $[s_0]$ defined in (\ref{Shor9}), we obtain
\begin{align}
\label{intrand}
s_k \bmod 2^{n-b-1} &= kr2^\alpha \bmod 2^{n-b-1} \cr
	&= (kr \bmod 2^{n-\alpha-b-1}) 2^\alpha,
\end{align}
where we assumed $s_0 = 0$ for analytical simplicity.
We note that $(kr \bmod 2^{n-\alpha-b-1})$ is a random integer variable in $k$
for $k$ an integer, which spans the entire integer space $0 \leq k \leq 2^{n-\alpha-b-1}-1$.
Now, we compute $\frac{\varphi}{\varphi_{max}}$, using (\ref{phimax}), (\ref{lbardef}), and 
(\ref{intrand}):
\begin{align}
\frac{\varphi(n,b,s_k,l)}{\varphi_{max}} &= \frac{\pi}{2^{n-1}} \frac{(s_k \bmod 2^{n-b-1}) 
	\times \bar{l}}{2\pi[2^{-b-1}(n-b) - 2^{-b} + 2^{-n}]} \cr
	&\approx \frac{\bar{l}}{n-b-2} \frac{kr \bmod 2^{n-\alpha-b-1}}{2^{n-\alpha-b-1}},
\end{align}
where we again dropped the small $2^{-n}$ term. Thus, we write
\begin{equation}
\label{randphi}
\varphi(n,b,s_k,l) \approx \frac{\bar{l} \varphi_{max}}{n-b-2} \bar{R}_k,
\end{equation}
where we used
\begin{equation}
\label{ratrand}
\bar{R}_k = \frac{kr \bmod 2^{n-\alpha-b-1}}{2^{n-\alpha-b-1}},
\end{equation}
which is a random variable in $k$ whose range is $[0,1)$.

We are now ready to evaluate (\ref{analyticalbasis}). Inserting
(\ref{randphi}) in (\ref{analyticalbasis}), we obtain
\begin{equation}
P(n,b) = \langle | \langle \exp \bigg(-i \bar{R}_k \frac{\varphi_{max}\bar{l}}{n-b-2} \bigg)
\rangle_k |^2 \rangle_j.
\end{equation}
Assuming that $\bar{R}_k$ is uniformly distributed in $[0,1)$, we turn the
$k$ average into an integral and obtain
\begin{equation}
\label{ARCintegral}
P(n,b) \approx \langle \Bigg| \int_{0}^{\eta} e^{-i\bar{R}} \frac{1}{\eta} d\bar{R} \Bigg|^2 \rangle_j,
\end{equation}
where we defined
\begin{equation}
\label{eta}
\eta = \frac{\bar{l} \varphi_{max}}{n-b-2}.
\end{equation}
Evaluation of (\ref{ARCintegral}) yields
\begin{equation}
\label{etaprob}
P(n,b) \approx \langle \frac{2}{\eta^2} [1-\cos(\eta)] \rangle_j.
\end{equation}
Since $\eta$ defined in (\ref{eta}) is small for $n < n_t$,
we Taylor-expand (\ref{etaprob}),
which results in
\begin{equation}
\label{etaprobtaylor}
P(n,b) \approx \langle \frac{2}{\eta^2} \bigg[1-\bigg(1-\frac{\eta^2}{2} + \frac{\eta^4}{24}\bigg)\bigg]
\rangle_j = 1-\frac{\langle \eta^2 \rangle_j}{12}.
\end{equation}
Inserting $\eta$ defined in (\ref{eta}) into (\ref{etaprobtaylor}), we obtain
\begin{equation}
\label{lbarprob}
P(n,b) \approx 1- \frac{\varphi_{max}^2 \langle \bar{l}^2 \rangle_j}{12(n-b-2)^2}.
\end{equation}
We compute $\langle \bar{l}^2 \rangle_j$ in the following way. Computing the 
average of the square of (\ref{lbardef}), we obtain
\begin{align}
\label{sqlinphiavg2}
\langle \varphi^2 \rangle_{kj} &= \frac{\pi^2}{2^{2n-2}} \langle ( s \bmod 2^{n-b-1} )^2 \rangle_k
\langle \bar{l}^2 \rangle_j \cr
&= \bigg( \frac{\pi^2}{3} \bigg) 2^{-2b} \langle \bar{l}^2 \rangle_j,
\end{align}
where we used the assumed independence of $s$ and $l$ of the random model.
According to (\ref{sqphiavg}), and to leading order in $x$ [defined in (\ref{xdef})], we have
\begin{equation}
\label{modsqlinphiavg2}
\langle \varphi^2 \rangle_{kj} \approx \bigg( \frac{\pi^2}{16} \bigg) 2^{-2b} (n-b-2)^2.
\end{equation}
Equating (\ref{sqlinphiavg2}) and (\ref{modsqlinphiavg2}), we obtain
\begin{equation}
\label{sqlbaravg}
\langle \bar{l}^2 \rangle_j = \frac{3}{16} (n-b-2)^2.
\end{equation}
Inserting (\ref{sqlbaravg}) into (\ref{lbarprob}), we obtain
\begin{equation}
P(n,b) \approx 1 - \frac{\varphi_{max}^2}{64} \approx \exp[-\varphi_{max}^2(n,b)/64].
\end{equation}
Compared with the numerical fit line (\ref{gaussfit}) [in particular equation (\ref{rawproblittle})],
this analytical result predicts the functional form of
the $b$-scaling exactly and the overall scaling factor within a factor of 2. 

\section{Comparison with the work of Fowler and Hollenberg}
\label{SectionFH}

\begin{figure}
\centering
\includegraphics[scale=1]{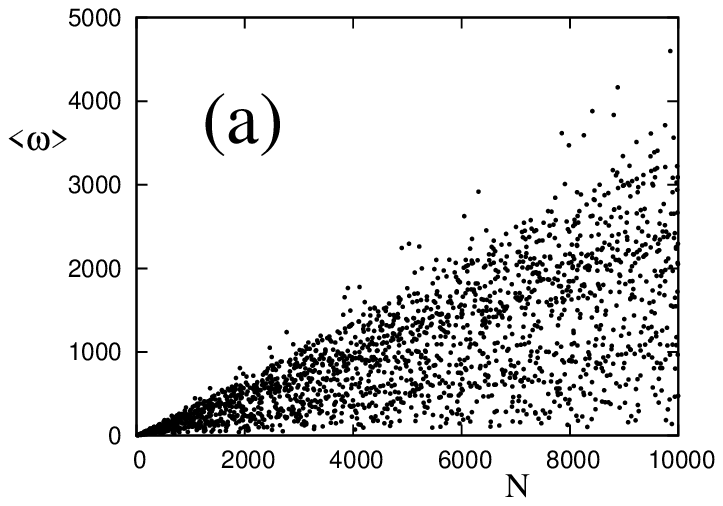} \\
\includegraphics[scale=1]{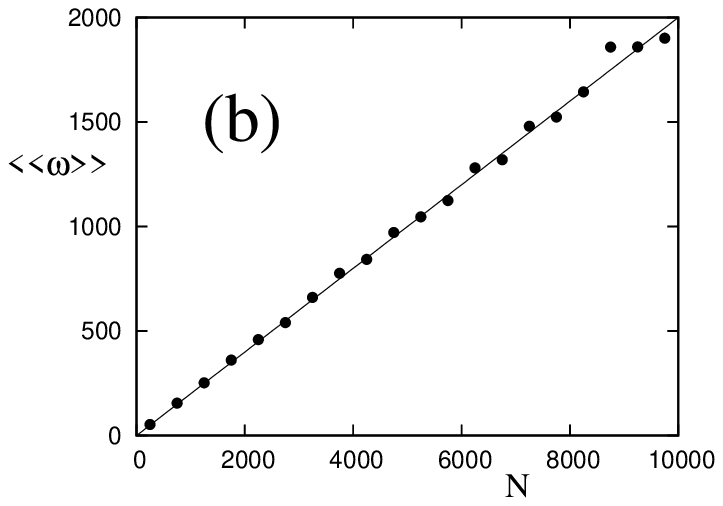}
\caption{\label{avgorderfigure} 
Average $\omega$ as a function of $N$. (a) Scatter plot of
$\langle \omega \rangle$ defined according to (\ref{avgomegadef});
(b) double averaged, binned
$\langle \langle \omega \rangle \rangle$ defined according to 
(\ref{doubleavgdef}).
  }
\end{figure}

Our work is closely related to the work of Fowler and Hollenberg \cite{FH}
(in the following abbreviated to FH). The purpose of this section is to discuss 
similarities and differences between the two approaches. The 
notation in \cite{FH} differs from ours. In order to avoid confusion, 
we translate the notation in \cite{FH} into our notation. 
As argued in \cite{FH} and here, because of the sensitivity of quantum gates 
to noise and decoherence, it is important to reduce the number of 
gates and gate operations as much as 
possible. This provides the motivation for studying the performance 
of Shor's algorithm as a function of bandwidth $b$ of the quantum 
Fourier transform, since a small $b$ results in substantial savings 
in gates to be implemented and gate operations to be executed. 
Both works conclude that for large $n$ 
the period-finding part of Shor's algorithm scales 
exponentially in $n$, $P(n,b)\sim 2^{-\xi_b n}$, where $\xi_b=\gamma 2^{-2b}$ 
and $\gamma$ a constant. FH quote $\gamma=2$; we find 
$\gamma=1.1$. 
Thus, while the research goals are the same, and the central results are 
similar, 
there are substantial differences in how the research programs are executed, and 
there are new findings in our work. Among the new findings is the 
existence of a non-exponential regime for small $n$ (see Sec.~\ref{SectionNR}),
analytical results for the non-exponential and exponential regimes 
(see Sec.~\ref{SectionAR}) and the existence of a provable bound 
for the maximal possible period $\omega$ of a given semiprime $N$ 
(see Appendix~\ref{AppendixB}). 
 
The main difference between \cite{FH} and our work concerns the choice of $\omega$ 
in the simulations. While in our work we simulate the period-finding part 
of Shor's algorithm 
for actual semiprimes $N$, and actual, associated $\omega$ 
values, FH use an effective $\omega=2+N/2$. Thus, our calculations 
are more realistic than those reported in \cite{FH} and check and complement the 
calculations in \cite{FH} under more realistic conditions. A first comment in this 
connection concerns the choice of FH's effective $\omega$ value. 
It was chosen as a good representative of $\omega$ values in Fig.~5 of 
\cite{FH}. However, the $\omega$ values in this figure extend up to $\omega=N$, 
which is more than twice larger than the maximal possible $\omega$, which is 
smaller than $N/2$ (see Appendix~\ref{AppendixB} for the proof). 
Therefore, rather than located in the middle of Fig.~5 of \cite{FH}, FH's 
effective $\omega$ actually lies beyond the allowed range of $\omega$. 
However, this is not expected to make any difference in the conclusions 
of \cite{FH}, since, as shown in Fig.~5 of \cite{FH}, according to the 
simulations reported in \cite{FH}, $P(n,b)$ exhibits flat 
plateaus in $\omega$. 

In this connection it may be interesting to present more information 
on the distribution of allowed $\omega$ values. In Fig.~\ref{avgorderfigure}~(a) 
we show the properly averaged $\omega$ values, 
\begin{equation}
\label{avgomegadef}
\langle \omega \rangle = 
\frac{1}{\varphi_E(N)} 
\sum_{j=1}^{a(N)} \nu(\omega_j) \omega_j 
\end{equation}
as a function of $N$ in the form of a scatter plot. 
The symbols in (\ref{avgomegadef}) have the same meaning as explained 
in connection with (\ref{NUMRES2}), i.e. 
$\varphi_E(N)$ is Euler's totient function, $a(N)$ is the 
number of $\omega$ values for given $N$, and 
$\nu(\omega)$ is the multiplicity of $\omega$. We see that 
$\langle\omega\rangle$ is a sensitive function of $N$ with a large 
spread over the entire allowed $\langle \omega \rangle$ range, i.e. 
$2\leq \langle \omega \rangle < N/2$. To make more sense of the 
raw $\langle\omega\rangle$ data, Fig.~\ref{avgorderfigure}~(b) 
shows a binned 
average of the $\langle \omega\rangle$ data 
in Fig.~\ref{avgorderfigure}~(a) defined as 
\begin{align}
\label{doubleavgdef}
\langle \langle \omega \rangle \rangle (N^{(i)})&= \frac{1}{\chi(N^{(i)}+250)-\chi(N^{(i)}-250)}
\sum_{\lambda=\chi(N^{(i)}-250)+1}^{\chi(N^{(i)}+250)} \langle \omega \rangle_\lambda, \cr
&N^{(i)} = 500\bigg(i-\frac{1}{2}\bigg),i=1,\ldots,20,
\end{align}
where $\chi(N)$ is the semiprime counting function and $\langle \omega \rangle_\lambda$
is the average $\omega$ [see (\ref{avgomegadef})] associated with the $\lambda$th semiprime.
Figure~\ref{avgorderfigure}~(b) shows that the twice averaged $\langle \langle \omega \rangle
\rangle$ 
are linear in $N$ with 
\begin{equation}
   \langle\langle \omega \rangle\rangle \approx N/5. 
\end{equation}
Therefore, according to Fig.~\ref{avgorderfigure}~(b), 
a representative $\omega$ value for a given $N$ 
is an allowed $\omega$ value in the vicinity of $N/5$. 

In contrast to our choice of a single $l$ state representing a Fourier peak, FH
choose two $l$ states to represent a Fourier peak, one to the left and one to 
the right of the position of the peak's maximum. This choice is more symmetrical than ours, 
but, because of the uniform response of all states under a Fourier peak 
(see Fig.~\ref{widthfigure} and the discussion in Sec.~\ref{SectionPM}),  
one representative is sufficient.

FH quote $\gamma_{FH}=2$ as a safe estimate, which is about 
a factor 2 larger than our, more optimistic, $\gamma=1.1$. On the 
basis of the data in Fig.~6 of \cite{FH} we computed the 
actual $\gamma_{FH}$ corresponding to the six panels of 
FH's Fig.~6, and obtained $\gamma_{FH}=0.5$ ($b=0$), 
1.85 ($b=1$), 1.83 ($b=2$), 1.79 ($b=3$), 1.78 ($b=4$), 
1.77 ($b=5$), 1.73 ($b=6$), and 1.57 ($b=7$). 
Discarding the $\gamma_{FH}$ value for $b=0$ (it is not 
generic, since it involves only H and M gates and no 
rotation gate), and the $\gamma_{FH}$ values for 
$b=6$ and $b=7$ (given the numerical 
range of the data, the exponential regime displayed 
in Fig.~6 of \cite{FH} is very short, resulting in uncertainty 
in the decay constant of an exponential fit),
the $\gamma_{FH}$ values are well characterized 
by $\gamma_{FH} \approx 1.8$, slightly more optimistic 
than the quoted $\gamma_{FH}=2$. What is interesting 
for us is that $\gamma_{FH}=1.8$ is already closer to 
our value of $\gamma=1.1$. 
 
Finally, what difference does it make for the performance 
of a quantum computer if $\gamma=2$ or $\gamma=1.1$? 
The answer depends on the performance level 
of the quantum computer. Since a factor 2 difference 
in $\gamma$ is the difference between performance and 
the square of the performance, a factor 2 difference in 
gamma has basically no effect if the quantum computer 
operates with close to 100\% performance, but has a {\it large} 
effect, if the quantum computer operates, e.g., on 
the 10\% level. 

Because of the critical need for quantum error correction and 
fault-tolerant operation \cite{WLH}, FH also present an error-tolerant, 
approximate construction of rotation gates, 
consisting of more fundamental elementary 
gates. In fact, each single-qubit rotation 
gate, as written in the quantum algorithm, may result in thousands 
of gates when decomposed. Unlike FH, we did not discuss the actual 
realization of gates, since, in this paper, we focus on 
the algorithmic aspects of Shor's algorithm, in particular on 
the scaling of the performance with $n$ and $b$. In any case, as shown by 
FH, the actual experimental realization of fault-tolerant gates 
may require large numbers of additional, ancillary gates 
and qubits, motivating and 
emphasizing the critical need to reduce 
required quantum resources as much as possible by optimizing 
the quantum algorithms. 
 
Given that error correction and fault-tolerant operation may 
introduce many additional auxiliary gates and qubits, what 
happens to our scaling laws in this case? Since our scaling 
laws depend on two parameters, $b$ and $n$, 
the answer has 
two parts. (i) Error correction will not affect the 
$b$ scaling, since the possibility of reducing the 
full quantum Fourier transform to a narrow-band 
quantum Fourier transform 
with bandwidth $b$ is an intrinsic property of the mathematical 
structure of the Fourier transform itself that has nothing 
to do with quantum error correction. In fact, under noisy 
conditions, it may not even be a good idea to increase the 
bandwidth of the quantum Fourier transform, because 
the algorithmic accuracy of the transform gained might be 
more than offset by the errors introduced by the additional 
gates that are now exposed to noise and decoherence. 
(ii) It is clear that each computational qubit in Shor's algorithm 
has to be protected with quantum circuits that consist of 
additional qubits. However, since the scaling laws derived in this paper
refer to the number $n$ of computational qubits,
our scaling laws remain unchanged.
 
Summarizing the discussion in this section,
we see our work as complementary to the pioneering work of FH, 
adding new insights, and confirming the major 
conclusions of FH, using an independent approach
based on period-finding simulations of actual semiprimes $N$,
supported by analytical results.

 
\section{Discussion}
\label{SectionDC}

An absolute limit of classical computing is reached 
when the physical requirements exceed the resources of 
the universe. According to this definition we can safely say 
that a classical computer, 
no matter its precise architecture, 
using the best currently available 
factoring algorithms, 
will never be able to factor a semiprime with 
5000 decimal digits or more. 
We see this in the following way. 
The best currently known algorithm for 
factoring large, ``hard'' semiprimes 
(more than ${\sim 130}$ decimal digits; no small 
factors) 
is the general number field sieve (GNFS) \cite{Pomerance}. 
It was recently used by Kleinjung {\it et al.}  \cite{Kleinjung} to 
factor the RSA challenge number 
RSA-768 (232 decimal digits). 
This factorization took 
the equivalence of 2000 years on a 2.2 GHz Opteron 
workstation \cite{Kleinjung}. The performance of 
the GNFS 
scales approximately as \cite{Pomerance} 
\begin{equation}
P(N) \sim \exp\left\{ 1.9 [\ln(N)]^{1/3} [\ln \ln (N)]^{2/3} \right\}, 
\label{DISC1}
\end{equation}
where $N$ is the semiprime to be factored. 
If we take the Kleinjung {\it et al.} factorization 
as the current, best benchmark, and estimate
an Opteron processor to consist of roughly $10^{25}$ 
particles, then we can factor a 232-decimal-digit 
semiprime with 
$2000\times 12\times 10^{25}\approx 2\times 10^{29}$ 
particles in the time span of a month. According 
to (\ref{DISC1}), then, 
in order to factor 
a 5000-decimal-digit number in the span of a month we 
need 
\begin{equation}
2\times 10^{29}\times P(10^{5000}) / P(10^{232}) \approx 
10^{89} 
\label{DISC2}
\end{equation}
particles. This exceeds the number of particles in the 
universe ($\approx 10^{80}$) by several orders of 
magnitude. Clearly, the factorization of a 
5000-decimal-digit semiprime is physically impossible 
to perform within a reasonable time ($\sim$ 1 month) 
on a classical computer. Even if we allow substantial 
progress in computer development, for instance replacing the 
current MOSFET transistors \cite{MOSFET} used in computer chips
by single-electron transistors \cite{Kastner} and increasing the 
clock-speed of a processor from 2.2 GHz to the optical 
regime of $\sim 10^{15}$ Hz, we gain only insignificantly. 
Therefore, in the absence of a breakthrough in the design of 
classical factoring algorithms, if we want to make any progress in factoring 
large numbers, we need a different computing paradigm. 
This is provided by switching from classical computing 
to quantum computing, i.e., running 
Shor's algorithm on a quantum computer. Instead of scaling 
(sub) exponentially, according to (\ref{DISC1}), Shor's 
algorithm scales $\sim O[(\ln N)^2(\ln \ln N)(\ln \ln \ln N)]$ \cite{Shor1}
and thus provides an 
exponential speed-up that allows us, in principle, to tackle semiprimes 
vastly in excess of $N = 10^{5000}$.
Obviously, for the practical implementation of powerful quantum
computers, any optimization of quantum algorithms is welcome.
Addressing this point, our paper shows that replacing the 
full quantum Fourier transform in Shor's algorithm with a 
narrow-band version incurs only a negligible performance penalty.
We also show how the performance of such a streamlined
version of Shor's algorithm scales with the number of qubits $n$.

In order to objectively characterize the performance of a 
quantum computer with $n$ qubits, 
equipped with a banded quantum Fourier transform 
of bandwidth $b$, 
we defined the performance measure 
$P(n,b,\omega)$ in Sec.~\ref{SectionPM} [see (\ref{measure})]. 
This measure was carefully chosen to 
accurately reflect the performance of the quantum computer 
in terms of the probability of a successful factorization, yet 
not excessively expensive to compute numerically and, most 
importantly, a convenient starting point for analytical computations. 
As shown in Secs.~\ref{SectionNR}~and~\ref{SectionAR}, 
our performance measure fulfills both goals. 
Although any given peak in the quantum Fourier transform 
contains several $l$ states with significant overlap with the 
Fourier peak, and 
useful for factorization in classical post-processing \cite{NC,Mermin}, our 
performance measure defined in  (\ref{measure}) 
is based only on a single $l$ state, i.e. the state $|l_j\rangle$ closest to 
the central maximum of the Fourier peak number $j$ [see (\ref{peakloc})].
This, no doubt, is 
convenient for analytical calculations, as successfully demonstrated 
in Sec.~\ref{SectionAR}, 
and for the following 
reason it is also justified. Numerically investigating the response of 
the Fourier peaks to a reduction of the bandwidth $b$, we found that 
the width of the Fourier peaks stays the same (about one state) while 
the height of the Fourier peaks is reduced. Thus, all $l$ states under 
a Fourier peak respond in unison to a change in $b$ (see Fig.~\ref{widthfigure}), and since 
the width of the Fourier peaks stays the same, the number of 
significant states in a peak is 
conserved, too. This means that a single state under the peak, 
for instance, the state with maximal overlap, accurately represents the 
response of any other state under the peak, in particular 
the states useful for factorization. Thus, 
summarizing our choice of performance measure, we may say that, 
of course, choosing all those states under a Fourier peak 
that are useful for factorization, would be best. However, this is 
computationally prohibitively expensive and not useful for analytical calculations. 
A proxy is necessary. Because of the uniform response of all 
states in a Fourier peak, this proxy is provided, e.g., by the state closest 
to the central peak, $| l_j \rangle$, and leads directly to our performance measure
$P(n,b)$ defined in (\ref{measure}). 

The exponential fit function in (\ref{expfit}) is 
shifted by 8 units in $n$. A possible explanation is the following. 
$n=8$ corresponds to $N=15$, the smallest odd semiprime. 
However, for $N=15$ all possible orders $\omega$ are 
powers of 2. Therefore, according to the discussion 
in Sec.~\ref{SectionPM}, Shor's algorithm 
performs perfectly in this case 
for all $b$. This means that 
$P(n=8,b,\omega)=1$ for all $b$, which is true independently 
of $b$ only if $\xi_b$ is multiplied with $n-8$ in 
the exponent of (\ref{expfit}). 

The largest RSA challenge number \cite{RSACHNUM} is 
RSA-2048. It has 2048 binary digits, which corresponds to 
617 decimal digits. Factoring this number on a quantum 
computer requires a minimum of 4096 qubits. 
As an illustrative example, let us assume 
that we factor this number on a 
quantum computer with $b=8$. Since no 
numerical simulation data are available in this 
very-large $n$ regime, we have to rely on our 
results (\ref{expfit}) and (\ref{gaussfit}) to 
estimate the performance of the quantum computer. 
Which of the two formulas to use depends on which 
regime, exponential or non-exponential, we are in. 
For $b=8$, and 
according to (\ref{crossformula}), the transition point 
$n_t$ for $b=8$ occurs at $n_t=20$.
Therefore, since $n \gg n_t$ in this case, we are sure 
that we are not in the non-exponential regime. 
However, how certain can we be that the exponential
law (\ref{expfit}) is valid all the way up to $n=4096$, when
we checked it numerically only up to $n\approx30$ (see Sec.~\ref{SectionNR})?
 
We answer this question in the following way.
The moment expansion (\ref{momentexpansion}) is 
certainly valid out to $n$ values for which our low-order Taylor expansion 
of $\exp(-i\varphi)$ is valid, i.e., for $\varphi < 1$. Since 
$\varphi < \varphi_{max}$, the safest estimate for the 
validity of (\ref{expfit}) is $n\lesssim 2^{b+1}/(2\pi)$, which is 
obtained from (\ref{phimax}) for $n\gg b$. For $b=8$ this implies 
$n<81$. This is already deeply in the $n$ regime where 
current numerical simulations cannot follow.
However, we can do better than that. 
The moment expansion (\ref{momentexpansion}), together with 
our numerical observation that the 4th order terms 
are given by $(\hat{\sigma}^2)^2/2$ shows that 
the relevant expansion parameter of (\ref{momentexpansion}) is not 
$\varphi$, but $\hat{\sigma}^2$, which is 
much smaller than $\varphi_{max}^2$. Therefore, 
we can safely assume exponential decay out to 
$n$ values for which $\hat{\sigma}^2<1$. According 
to (\ref{variance}), then, this yields the estimate 
$n<12\times 2^{2b}/\pi^2$, which amounts to 
$n<79682$ for $b=8$, much larger than $n=4096$ required 
for the factorization of RSA-2048. We conclude that, 
for $b=8$, we may safely use the exponential 
law (\ref{expfit}) to estimate the performance of the 
quantum computer. Therefore, using $n=4096$ and $b=8$ 
in (\ref{expfit}), we obtain $P(n,b)=0.954$, i.e. a quantum computer with 
a bandwidth of only $b=8$ can factor the RSA challenge number 
RSA-2048 with a performance of better than 95\%. 
If we increase $b=8$ by only one unit to $b=9$, the performance 
increases to 98\%. 

Concluding this section, we briefly discuss the 
paper by Barenco {\it et al.} 
\cite{BEST}, which also investigates the 
effect of the banded quantum Fourier transform on 
the performance of the period-finding part of 
Shor's algorithm. In fact, their performance measure $Q$, 
based on the probability of obtaining 
an $|l\rangle$-state closest to $2^n/\omega$, is, 
up to normalization, identical with our performance measure. 
However, the main focus of \cite{BEST} is the 
effect of decoherence on $Q$ and, similar to the 
work of Fowler and Hollenberg \cite{FH}, 
Barenco {\it et al.} do not use factoring of actual 
semiprimes $N$ in their numerical simulations. 
Finally, the analytical performance estimates in 
\cite{BEST} require $b>\log_2(n)+2$, which, 
for $b=8$, implies $n<64$. Therefore, for small 
$b\lesssim 8$, 
the analytical 
formulas of \cite{BEST} are not applicable 
to the performance of a quantum computer 
in the technically and commercially interesting 
small-$b$, large-$n$ regime with $n\gtrsim 4000$. 
              
 
\section{Summary and Conclusions}
\label{SectionSC}
Given that quantum computers are difficult to build,
any advance in the optimization of quantum algorithms
is welcome. Accordingly, in this paper, we investigated
the performance of Shor's algorithm equipped with
a banded quantum Fourier transform. Our predictions 
are based on the following five substantial advances.
\begin{enumerate}
\item Properly $\omega$-averaged numerical simulations of 
      factoring actual semiprimes $N$ for qubit numbers 
      ranging from $n=9$ to $n=33$, yielding the 
      numerical performance estimates (\ref{expfit})
      in the large-$n$ regime 
      and (\ref{gaussfit}) in the small-$n$ regime. 
\item Analytical and numerical justification of the separation 
          of the $k$ and $j$ sums in the definition of the performance 
          measure as the foundation of analytical computations 
          of the performance measure in the large-$n$ and 
          small-$n$ regimes. It is shown that both separations 
          are exponentially 
          accurate, with exponential improvement of accuracy 
          for increasing bandwidth $b$ of the quantum Fourier transform. 
\item Analytical computation of the performance measure in 
      the exponential, high-$n$ regime, which predicts the $2^{-2b}$
      scaling exactly and the prefactor in $\xi_b$ within $10\%$ of the
      numerical result (\ref{expfit}).
\item Analytical computation of the performance measure 
      in the small-$n$ regime, which predicts the functional form 
      of the performance measure accurately and provides 
      a reasonable estimate of a single, overall scaling factor. 
\item Analytical formula (\ref{crossformula}) for the cross-over points 
      $n_t$ that mark the transition from the 
      non-exponential regime to the exponential regime 
      of quantum computer performance. For given 
      bandwidth $b$ and number of 
      qubits $n$, this allows a quick, 
      accurate, and convenient 
      decision of whether the resulting finite-bandwidth 
      quantum computer 
      is working in the exponential or 
      non-exponential regime. 
\end{enumerate}
In addition, in Appendix~\ref{AppendixA}, 
we prove the 
existence and uniqueness of an order-2 seed 
for any semiprime $N$, which, 
in Appendix~\ref{AppendixB}, 
is used to 
prove that the maximal possible order $\omega$ 
of a seed is less than $N/2$ (see Figs.~\ref{avgorderfigure}~and~\ref{maxorderfigure}).
The maximally allowed $\omega$ is smaller than
the effective, representative $\omega$ chosen in \cite{FH}.
However, due to the insensitivity of the results in \cite{FH}
with respect to the chosen $\omega$ (see Fig.~5 of \cite{FH}),
this fact is not expected to change the results predicted in \cite{FH}.
Lastly, we investigate the statistical properties of an inverse factor
of $\omega$ in Appendix~\ref{AppendixC}.
 
In our opinion, and based on the numerical and analytical 
results presented in this paper, we conclude that 
the period-finding part of Shor's algorithm equipped with 
a banded quantum Fourier transform of bandwidth $b$ is 
now essentially understood. 
However, period-finding is not the most demanding part
of Shor's algorithm to implement.
This distinction is 
reserved for the $f$-mapping part of Shor's algorithm (the modular exponentiation part),
which feeds register $II$ with $f(s)$ values 
(see Sec.~\ref{SectionShor}) and, compared with the period-finding 
part of Shor's algorithm, requires vastly 
more quantum resources to implement 
 \cite{EkJo,Shep,Vedral,Meter}. 
Therefore, attention now has to be directed toward optimizing 
the $f$-mapping part of Shor's algorithm. 
%
              
\appendix

\section{Existence and Uniqueness of an element
                 of Order 2} 
                 \label{AppendixA}
In support of the result that the probability of encountering 
a seed with a small order is small, we provide here 
a proof that there is one and only one seed $x$ of order $2$ 
for any semi-prime $N=pq$, where $p\neq q$ are primes 
larger than 2. A seed is any positive integer, larger than 1, 
that is relatively prime to $N$. Let us collect all 
possible seeds $x_j$, $j=1,\ldots,L-1$, including 
the unit 1, into a set 
$G_N = \{1,x_1,x_2,\ldots,x_{L-1}\}$. This way, $G_N$ forms a multiplicative 
group modulo $N$ \cite{Algebra} containing $L$ elements. 
 
The computation of $L$ is straightforward. There are at most 
$N-1$ numbers that are 
relatively prime to $N=pq$. 
(By definition, the unit element 1 is relatively prime to $N$ 
\cite{Jacobson}, but $N$ is not.) 
However, 
$p-1$ of these numbers contain a factor $q$ and $q-1$ of 
these numbers contain a factor $p$, and these numbers 
are all different. 
Therefore, there are 
$L=(N-1)-(p-1)-(q-1)=N-p-q+1$ group elements. Since 
$N$, $p$, and $q$ are odd, $L$ is even. At this point 
we cite a well-known theorem of elementary algebra that 
states that each group with an even number of elements 
has at least one element that is different from the 
unit element and is of order two \cite{Jacobson}. 
Applied to our group $G_N$ this means that there exists 
at least one seed $x\neq 1$ with $x^2=1$ modulo $N$, 
i.e. a seed of order 2. 

At this point it is important to observe that if there is 
a seed $x$ with $x^2\mod N=1$, then there is a mirror 
seed $z=N-x$, which is also of order 2, since 
$z^2\mod N=(N^2-2Nx+x^2)\mod N=x^2\mod N=1$. 
Therefore, without restriction of generality, we 
will restrict ourselves to the range of seeds 
smaller than $N/2$ and prove 
that there is only
one $x<N/2$ with $x^2\mod N=1$, where $N = pq$. 
 
We already proved that there is at least one
$x$ with 
\begin{equation}
x^2 \mod N = 1. 
\label{ExUni0a} 
\end{equation}
Without restriction of generality, 
we can choose this $x$ to be smaller than $N/2$, since, 
if it is larger than $N/2$, its mirror will be smaller than $N/2$. 
Assume that there exists another seed of order 2, $y<N/2$,
with $y>x$ (no restriction of generality) and 
\begin{equation}
y^2 \mod N = 1.
\label{ExUni0b}
\end{equation}

Since $x^2 \mod N = 1$ and $y^2 \mod N = 1$, we have
\begin{equation}
(y^2 - x^2) \mod N = (y-x) (y+x) \mod N = 0.
\label{ExUni0c}
\end{equation}
This equation holds if either 
(i) at least one of the factors is divisible by $N$ or
(ii) $(y-x)$ contains $p$ and $(y+x)$ contains $q$, or
          vice versa.
However, case (i) is impossible:
Since both $x$ and $y$
are smaller than $N/2$, $(y+x)<N$ is, therefore, never divisible 
by $N$. For the same reason $(y-x)$ is divisible 
by $N$ only if $(y-x)=0$, which is excluded, 
since, according to assumption, $y\neq x$. 
This leaves case (ii).
 
Since $x^2 \mod N = 1$, we have
$(x-1) (x+1) \mod N = 0$. Since
$(x-1) < N$ and $(x+1) < N$, for any $N>2$,
neither factor is divisible by $N$ and 
the product is divisible by $N$ only if 
$(x-1)$ is a multiple of $p$ and $(x+1)$ is a multiple of $q$. 
There is no restriction of generality here, since 
which factor of the product is divisible by which factor of 
$N$ ($p$ or $q$) 
is merely a matter of properly labeling the factors of $N$. 
So, let us write:
\begin{align}
     x-1 &= \lambda p ,
     \label{ExUni1'}
     \\
     x+1 &= \mu q , 
     \label{ExUni1}
\end{align}
where $\lambda$ and $\mu$ are positive integers. 
We observe immediately that 
$\lambda$ cannot contain a factor $q$, since 
otherwise $(x-1)$ would be divisible by $N$.
In the same way we reason that $\mu$ cannot 
contain a factor $p$. We record this observation as 
\begin{align}
  \lambda \mod q &\neq 0,
  \label{ExUni2'}
  \\
  \mu \mod p &\neq 0. 
  \label{ExUni2}
\end{align}
We also have $y^2 \mod N = 1$, i.e. 
$(y-1) (y+1) \mod N = 0$, 
which now implies two
possibilities, 
since in (\ref{ExUni1'}) and (\ref{ExUni1}) we already chose the naming convention 
for the two factors $p$ and $q$ of $N$. The two cases are: 
\begin{align}
 &\text {(A)\ \ \   $(y-1)$ is a multiple of $p$, $(y+1$) is a multiple of $q$} 
   \\ 
     &\text{(B)\ \ \   $(y-1)$ is a multiple of $q$, $(y+1)$ is a multiple of $p$.} 
\label{ExUni3}
\end{align}
Let us look at case (A) first. Let us write:
\begin{align}
     (y-1) &= \alpha p,
     \label{ExUni4'}
     \\ 
     (y+1) &= \beta q.
     \label{ExUni4} 
\end{align} 
In analogy with the reasoning that led us to (\ref{ExUni2'}) and (\ref{ExUni2}) we have 
\begin{align}
  \alpha \mod q &\neq 0,
  \label{ExUni5'}
  \\
  \beta \mod p &\neq 0. 
  \label{ExUni5}
\end{align}
Then, because of $x,y<N/2$, (\ref{ExUni0c}), and the 
discussion following (\ref{ExUni0c}), we need 
to prove 
that either $(y-x)$ contains a factor $p$ and $(y+x)$ a 
factor $q$ or vice versa. We write:
\begin{equation}
     y+x = (y-1) + (x+1) = \alpha p + \mu q.
\label{ExUni7} 
\end{equation} 
But since $\alpha$ is not divisible by $q$ [see (\ref{ExUni5'})] 
and $\mu$ is not divisible by $p$ [see (\ref{ExUni2})], 
$(y+x)$ is neither divisible by $p$ nor by $q$.
Therefore, case (A) leads to a contradiction, which implies that 
according to case (A) a second order-2 seed $y\neq x$ does not exist. 
 
Let us now look at case (B). Let us write: 
\begin{align}
     (y-1) &= \gamma q,    \\ 
     (y+1) &= \nu p,  
\label{ExUni8}
\end{align}
where, again, 
in analogy with the reasoning that led us to (\ref{ExUni2'}) and (\ref{ExUni2}), we have 
\begin{align}
  \gamma \mod p &\neq 0,
  \label{ExUni9'}
  \\
  \nu \mod q &\neq 0. 
  \label{ExUni9}
\end{align}
Then:
\begin{equation}
     y-x = (y-1) - (x-1) = \gamma q - \lambda p,
\label{ExUni10}
\end{equation}
which, because of (\ref{ExUni9'}) and (\ref{ExUni9}) is neither divisible by $p$ nor by $q$.
Therefore, case (B), too, leads to a contradiction.
 
As a result, we obtain that the existence of
an additional order-2 seed $y\neq x$, $y<N/2$ is impossible. 
Therefore, $x$ is the unique order-2 seed with $x < N/2$. 
This means that for any given semi-prime $N=pq$, there are exactly two 
order-2 seeds, $x<N/2$ and its mirror $N-x>N/2$. 
              
 
\section{Maximal Order} 
\label{AppendixB}

In connection with Shor's algorithm, 
for a given semi-prime $N$, we consider 
seeds $x$ with an {\it even} order $\omega=2\Omega$, 
where $\Omega\geq 1$ is a positive integer. The purpose 
of this section is to show that the largest possible even $\omega$ 
is smaller than $N/2$. 
 
A seed $x$, 
$1\leq x<N$ is a positive integer, relatively prime to $N=pq$, where 
$p\neq q$ are prime numbers larger than 2. As discussed in Appendix~\ref{AppendixA}, 
the set of seeds $x$ forms a group $G_N$ with 
\begin{equation}
|G_N| = N-p-q-1 = (p-1)(q-1) 
\label{SPO1}
\end{equation}
elements. We note that, according to (\ref{SPO1}), $|G_N|$ is 
divisible by 4, a fact which will become relevant below. 
If $x$ is relatively prime to $N$, so is $N-x$. Therefore, 
if $x$ is a seed, so is $N-x$, which implies (i) a symmetry of 
seeds with respect to $N/2$ and (ii) that there is an even number 
of seeds. We use (i) to define a set $\hat G_N$, consisting 
of elements $\hat x=(x,N-x)$, where $x$ and $N-x$ are identified. 
The set $\hat G_N$ forms a group. This is so, since $\hat G_N$ contains 
the unit element $\hat 1=(1,N-1)$, the product $\hat x \hat y$ of two elements of 
$\hat G_N$ is again in $\hat G_N$, and with each $\hat x$, we also find 
its inverse $(\hat x)^{-1}$ in $\hat G_N$. Because of (i) the 
group $\hat G_N$ has 
\begin{equation}
|\hat G_N| = |G_N| / 2
\label{SPO1a}
\end{equation}
elements. 

Let us form the set $G_N^*$ that contains the 
squares of $x$ modulo $N$. Since 
$G_N^*$ contains 
the unit element 1, and since with each $x^2$ and $y^2$ in 
$G_N^*$, the product 
\begin{equation}
(x^2) (y^2)\mod N=(xy)^2 \mod N 
\label{SPO2}
\end{equation}
is also in $G_N^*$, 
and since with each $x^2$ we also 
find its inverse 
\begin{equation}
(x^2)^{-1}\mod N=(x^{-1})^2\mod N
\label{SPO3}
\end{equation}
in $G_N^*$, the set $G_N^*$ is a group. 
In the same way we form the set $\hat G_N^*$ 
from the squares of $\hat x$ in $\hat G_N$. 
Because of the definition of 
$\hat G_N$, identifying $x$ and $N-x$, and 
because of 
\begin{equation}
(N-x)^2 \mod N = x^2 \mod N,
\label{SPO4}
\end{equation}
which shows that the squares of $x$ and $N-x$ are 
identical, the groups $G_N^*$ and $\hat G_N^*$ have 
the same number of elements. In addition, as is 
easily verified, 
the groups $G_N^*$ and $\hat G_N^*$ are isomorphic, 
which implies that the order of an element 
in $\hat G_N^*$ is the same as the order of an 
element in $G_N^*$. Let us denote the number of 
elements in these two groups by 
\begin{equation}
|G_N^*| = |\hat G_N^*| = M. 
\label{SPO5}
\end{equation}
Then, 
because of (\ref{SPO1a}), and because 
$\hat G_N^*$ is a subgroup of $\hat G_N$, 
we have that 
\begin{equation} 
\text{$M=|\hat G_N^*|$ divides $|\hat G_N| = |G_N|/2$}. 
\label{SPO6}
\end{equation}
One possibility is $M=|G_N|/2$. However, 
since the group $\hat G_N^*$ of squares is a subgroup of 
$\hat G_N$, $M=|G_N|/2$ is possible only if there 
are as many squares $\hat x^2$ in $\hat G_N^*$ 
as there are elements $\hat x$ 
in $\hat G_N$. However, because of the 
existence of a non-trivial order-2 element $\hat a$ 
(see Appendix~\ref{AppendixA}), this is impossible, since 
both $\hat 1^2=\hat 1$ and $\hat a^2=\hat 1$, 
which immediately implies $M<|G_N|/2$. 
Therefore, the largest possible $M$ that divides $|G_N|/2$ 
(an even number) is $|G_N|/4$, which implies 
\begin{equation}
M \leq |G_N|/4 . 
\label{SPO7}
\end{equation}
According to Euler's totient theorem \cite{Jacobson}, we have 
for any $\hat x^2$ in $\hat G_N^*$: 
\begin{equation}
(\hat x^2)^M = \hat 1,
\label{SPO8}
\end{equation}
which implies that the order of any element $\hat x^2$ in 
$\hat G_N^*$ is at most $M=|G_N|/4$. Because of the isomorphism 
between $\hat G_N^*$ and $G_N^*$, this implies that the order 
of any $x^2$ in $G_N^*$ is at most $|G_N|/4$. This, finally, implies 
that the order of any element $x$ in $G_N$ is at most $|G_N|/2$, i.e. 
\begin{equation}
\omega \leq |G_N|/2 < N/2.
\label{SPO9}
\end{equation}
%
\begin{figure}
\centering
\includegraphics[scale=1]{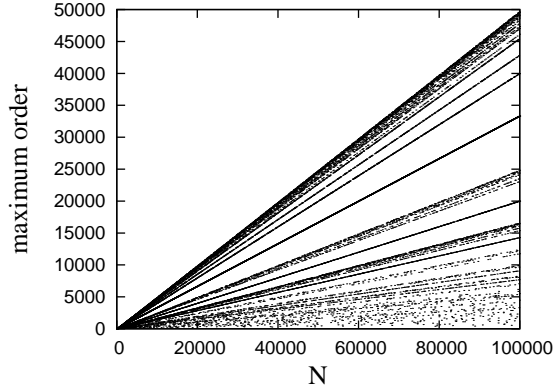}
\caption{\label{maxorderfigure} 
Maximal possible orders $\omega$ (maximum order)
computed and displayed for each $N$ in a complete
list of semiprimes in the interval $0 < N < 10^{5}$.
Apparently, the maximal possible order never exceeds $N/2$,
a fact proved in the text. 
}
\end{figure}
We note that since an essential element of the proof is to consider
the group of squares of $x$, the proof indeed applies only to {\it even} $\omega$.
An illustration of (\ref{SPO9}) is provided in Fig.~\ref{maxorderfigure},
which shows the maximum even orders of all semiprimes $N$ 
ranging up to $N=100000$. The figure illlustrates
(i) that the maximal order is indeed smaller than $N/2$ and
(ii) that the maximal order of a given semiprime $N$ is
not always close to $N/2$ but still has to divide the group order.
Therefore, in addition to the line $\sim N/2$, we also see the lines
corresponding to $\sim N/4$, $\sim N/6$, etc..

\section{$1/r$ average}
\label{AppendixC}
\begin{figure}
\centering
\includegraphics[scale=1]{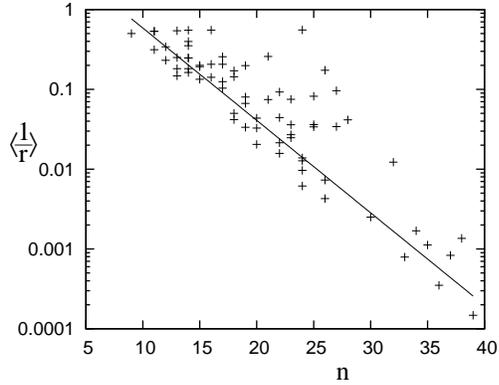}
\caption{\label{invrfigure} 
The fraction $\langle \frac{1}{r} \rangle$ as a function of $n$ for several semiprimes. 
The fit line (solid line) is the function $\langle \frac{1}{r} \rangle =2^{-(n-8)/2.6}$.
}
\end{figure}

For the analytical formula (\ref{rawproblittle}), we need the average $\langle \frac{1}{r} \rangle$
of $1/r$ as a function of $n$, where $r$ is defined in (\ref{orderdecomp}). We computed it
in the following way. 
First, we computed all possible orders, $\omega_j$, of a given semiprime $N$ with
their associated multiplicities, $\nu(\omega_j)$. 
Then, we extracted the odd part of the obtained orders, $r$, as defined in (\ref{orderdecomp}).
Denoting the odd part of a specific order $\omega_j$ by $r_j$, in analogy with
(\ref{NUMRES2}) and (\ref{avgomegadef}), we obtain
\begin{equation}
\label{avginvr}
\langle \frac{1}{r} \rangle = \frac{1}{\varphi_E (N)} \sum_{j=1}^{a(N)} \nu(\omega_j) \frac{1}{r_j},
\end{equation}
where the symbols in (\ref{avginvr}) share the same definition as shown in (\ref{NUMRES2})
and (\ref{avgomegadef}), i.e. $\varphi_E (N)$ is Euler's totient function and $a(N)$ is the number
of orders for given $N$.
Figure~\ref{invrfigure} shows the computed $\langle \frac{1}{r} \rangle$ according to (\ref{avginvr})
as a function of $n$, the number of qubits needed for a reliable determination of the order
as described in connection with (\ref{NUMRES1}). By graphically extracting 
the $n$-dependence of $\langle \frac{1}{r} \rangle$ using the fit line in Fig.~\ref{invrfigure}, 
we find
\begin{equation}
\langle \frac{1}{r} \rangle = 2^{-(n-8)/2.6}.
\end{equation}

\end{document}